# REHYBRIDIZATION DYNAMICS INTO THE PERICYCLIC MINIMUM OF AN ELECTROCYCLIC REACTION IMAGED IN REAL-TIME

Y. Liu,[1,2,†] D. M. Sanchez,[1,3,†,‡], M. R. Ware,[1] E. G. Champenois,[1] J. Yang,[1,4,5] J. P. F. Nunes,[6,7] A. Attar,[4] M. Centurion,[6] J. P. Cryan,[1] R. Forbes,[4] K. Hegazy,[1] M. C. Hoffmann,[4] F. Ji,[4] M.-F. Lin,[4] D. Luo,[4] S. K. Saha,[6] X. Shen,[4] X. J. Wang,[4,] T. J. Martínez,[1,3,*] & T. J. A. Wolf[1,*]

[1]Stanford PULSE Institute, SLAC National Accelerator Laboratory, Menlo Park, USA.

[2]Department of Physics and Astronomy, Stony Brook University, Stony Brook, NY, USA.

[3]Department of Chemistry, Stanford University, Stanford, USA.

[4]SLAC National Accelerator Laboratory, Menlo Park, USA.

[5]Center of Basic Molecular Science, Department of Chemistry, Tsinghua University, Beijing, China.

[6]Department of Physics and Astronomy, University of Nebraska-Lincoln, Lincoln, USA.

[7]Diamond Light Source, Harwell Science Campus, Didcot, UK

*Corresponding authors. Emails: thomas.wolf@slac.stanford.edu, toddjmartinez@gmail.com

†These authors contributed equally to this work.

‡Present address: Design Physics Division, Lawrence Livermore National Laboratory, Livermore, USA.

**Abstract:**

Electrocyclic reactions are characterized by the concerted formation and cleavage of both σ and π bonds through a cyclic structure. This structure is known as a pericyclic transition state for thermal reactions and a pericyclic minimum in the excited state for photochemical reactions. However, the structure of the pericyclic geometry has yet to be observed experimentally. We use a combination of ultrafast electron diffraction and excited state wavepacket simulations to image structural dynamics through the pericyclic minimum of a photochemical electrocyclic ring-opening reaction in the molecule α-terpinene. The structural motion into the pericyclic minimum is dominated by rehybridization of two carbon atoms, which is required for the transformation from two to three conjugated π bonds. The σ bond dissociation largely happens after internal conversion from the pericyclic minimum to the electronic ground state. These findings may be transferrable to electrocyclic reactions in general.

**One-Sentence Summary:** Rehybridization dynamics into the pericyclic minimum of an electrocyclic reaction imaged by ultrafast electron diffraction and ab initio multiple spawning simulations.

**Main Text**

Electrocyclic reactions, a subgroup of the class of pericyclic reactions, are important synthetic tools in organic chemistry due to their stereospecificity, i.e. their ability to control the reaction outcome through the stereochemistry of the reactant. Moreover, they are relevant to biological processes such as the biosynthesis of vitamin D in human skin.[1] Their stereospecificity originates from the concerted rearrangement of π and σ electrons through a single, cyclic critical geometry simultaneously making and breaking multiple bonds.[2] These simultaneous rearrangements require the involved π and σ molecular orbitals to exhibit a similar orientation to enable their interaction, which requires the rehybridization of some of the involved atoms. However, the detailed structure of the critical geometries of electrocyclic and pericyclic reactions have never been directly characterized experimentally due to the lack of methods with suitable sensitivity.

The concerted bond rearrangement can be illustrated by electrocyclic ring-opening in 1,3-cyclohexadiene (CHD)-like molecules. The two electrons participating in the σ bond, which is broken during the reaction (($C_3$-$C_4$), see **Fig. 1** for the CHD-derivative α-terpinene), and the 4 π-electrons of a system of two conjugated double bonds (($C_1$=$C_6$) and ($C_9$=$C_5$)) undergo a concerted transformation into a set of three conjugated double bonds. The positions of these double bonds change (($C_3$=$C_1$), ($C_6$=$C_9$), and ($C_5$=$C_4$)) during this process, resulting in a change in hybridization of the ($C_3$) and ($C_4$) carbon atoms from $sp^3$ (tetrahedral coordination) to $sp^2$ (planar coordination).

Electrocyclic reactions can proceed through thermal or photochemical pathways with opposite stereospecificity. The famous Woodward-Hoffmann rules predict this behavior based on the symmetry of the involved molecular orbitals and the structural motion from reactant through the critical geometry to the product.[2] For thermal electrocyclic reactions in the electronic ground state, the critical geometry is referred to as the pericyclic transition state. Due to their short lifetimes, transition states in general can rarely be experimentally characterized, with very few special exceptions.[3–6] Photochemical electrocyclic reactions take place on ultrafast timescales and are enabled by nonadiabatic dynamics through a conical intersection (CI) connecting the lowest excited state ($S_1$) to the ground state ($S_0$, see **Fig. 1**).[7,8] Here, the critical geometry represents a minimum in $S_1$ close to the CI, known as the pericyclic minimum.[9–11]

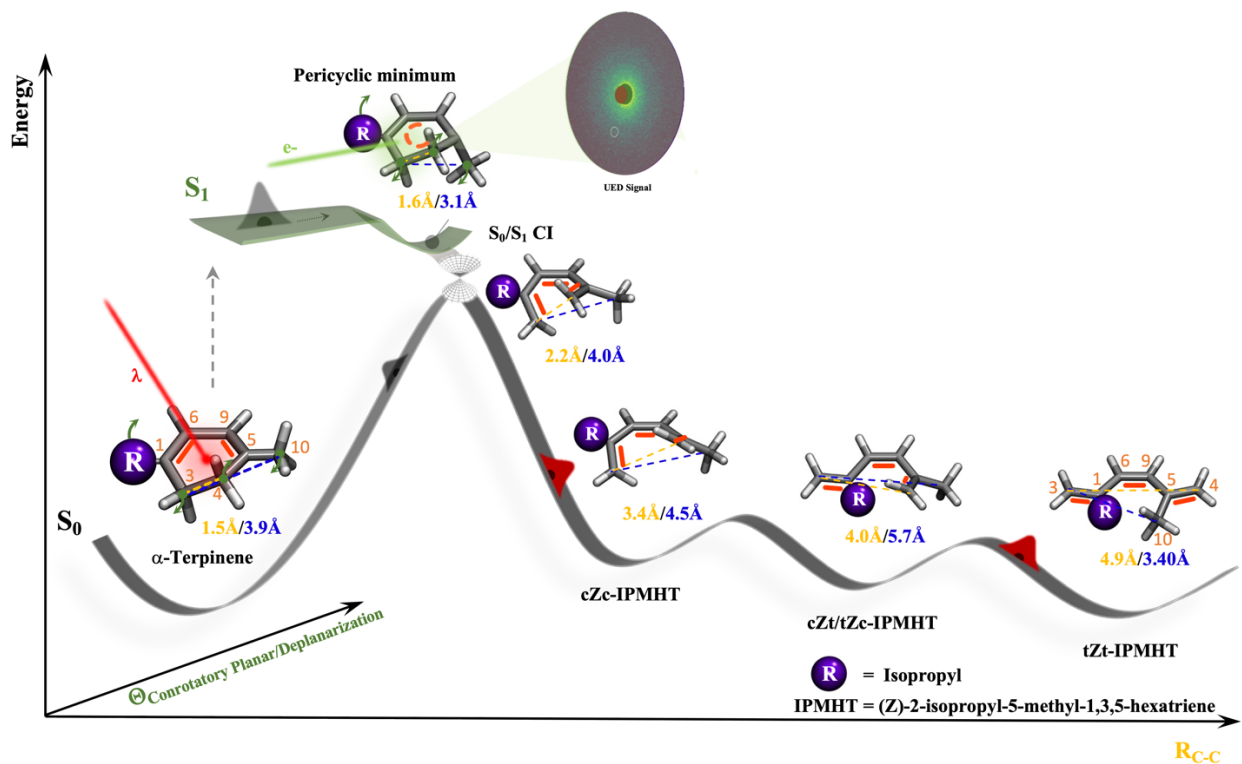

**Figure 1. Schematic description of the observed electrocyclic ring-opening dynamics of α-terpinene.** After photoexcitation to the first excited state ($S_1$), the molecule relaxes along a coordinate representing deplanarization with respect to the reactant double bond positions and planarization with respect to the product double bond positions into the pericyclic minimum. The pericyclic minimum is close to, but separated by a shallow barrier from a conical intersection ($S_0/S_1$ CI) with the electronic ground state ($S_0$). Population which relaxes through the CI either returns to the $S_0$ reactant minimum or evolves along a carbon-carbon bond dissociation coordinate $R_{C-C}$ into three $S_0$ minima representing different photoproduct isomers. Visualizations of representative structures along the reaction coordinate are shown together with specific carbon-carbon distances. Both the structures and the distances are extracted from the simulations. The carbon numbering used in the text is shown in orange. The double bond positions are highlighted in the structure visualizations as red bars.

The Woodward-Hoffmann rules state that the stereospecificity of the photochemical reaction pathway (which we confirmed in a previous study[12]) is guaranteed by the planarization during the rehybridization of the ($C_3$) and ($C_4$) $CH_2$ groups taking place in a conrotatory fashion, i.e., by rotation of the $CH_2$ groups in the same clockwise or counterclockwise direction. Thus, the pericyclic minimum geometry of CHD represents a critical snapshot of all the above mentioned simultaneous structural rearrangements: σ bond breaking, rehybridization, π bond alternation, and conrotatory motion.

In a recent, seminal study, the electronic structure changes that take place during relaxation of $S_1$-excited CHD from the reactant equilibrium geometry to the pericyclic minimum were visualized using time-resolved X-ray absorption spectroscopy.[13] Our present results provide complementary information about the nuclear structure changes during relaxation into the pericyclic minimum using a combination of megaelectronvolt ultrafast electron diffraction and *ab initio* multiple spawning simulations.[12,14–16] The high structural sensitivity of our methodology provides unprecedented access to the structural details of this photochemical "transition state" and provides a new perspective on the origins of the reaction's stereospecificity.

We study the α-terpinene (αTP) molecule, which differs from CHD by the addition of methyl and isopropyl substituents (see **Figs. 1** and **2**). The photochemical ring-opening of αTP has previously been studied with time-resolved spectroscopy. However, these were only indirectly sensitive to the structural dynamics of the molecule.[7,17–21] According to our experimental and simulation results, the presence of the substituents in αTP does not qualitatively alter the photochemical dynamics in comparison to CHD apart from a slight overall slowing of the dynamics.[15] Hydrogen atoms are difficult to track with electron diffraction because of their weak scattering. The addition of the methyl and isopropyl substituents introduces carbon atoms which act as "reporter" atoms by adding the stronger signatures of carbon-carbon bond distance changes to our experimental observable, time-dependent atomic pair distribution functions (PDFs). These signatures were missing in previous studies of the structural dynamics of CHD.[15,22] The "reporter" signatures provide direct evidence for a substantial part of the rehybridization and the conrotatory motion to take place in the $S_1$ state of αTP prior to internal conversion to $S_0$, which triggers σ-bond dissociation.

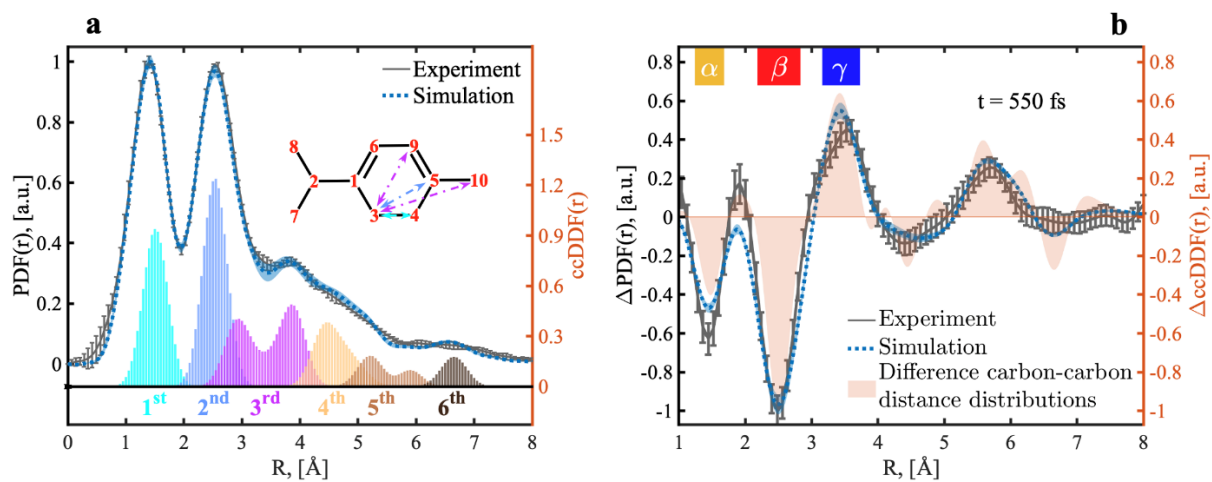

**Figure 2. Experimental and simulated structural information of αTP.** The line plots in panel **a** show both the simulated and experimental pair distribution functions (PDFs) of the molecule in the ground state. The histograms below the PDFs represent carbon-carbon distance distributions (ccDDF) based on the initial geometries of our *ab initio* multiple spawning simulations separated and color-coded with respect to carbon coordination spheres. The inset of panel **a** shows the labeling of the carbon atoms of αTP as used in the text. Additionally, representative distances for the first three coordination spheres are marked by color-coded arrows. Panel **b** shows experimental and simulated difference PDF (ΔPDF) at a pump-probe delay of 550 fs. The light-orange-colored area-plot indicates the total difference carbon-carbon distance distribution function (ΔccDDF) from all the carbon coordination spheres. Three regions are labeled as α, β, and γ. Uncertainties derived from bootstrapping analyses are shown as error bars (experiment) and shaded areas (simulation).

**Results**

In **Fig. 2a**, we show the experimental atomic pair distribution function (PDF) obtained by real-space transformation of static diffraction patterns from gas phase αTP. For comparison, we plot a simulated PDF of the ground state structure of αTP. The simulation is based on a diffraction signal obtained from the initial conditions of our excited state wavepacket simulations. (see **Methods** for further details). Experimental and simulated PDFs are in quantitative agreement. The main contributors to the PDFs are carbon-carbon distances, although they also contain weak signatures from C-H and H-H distances, which we will neglect in the following discussion (see

**Supplementary Note 1**). The structural information contained in the PDFs is conveniently presented in the framework of carbon-carbon coordination spheres. The carbon-carbon distance distributions extracted from the simulations are additionally shown color-coded with respect to the coordination spheres in **Fig. 2a**. Representative distances from the first three coordination spheres are shown as color-coded arrows in the inset of **Fig. 2a**.

The peak at 1.4 Å in the PDFs can be associated with the first coordination sphere (cyan) representing carbon-carbon bond distances. The largest contributions to the peak at 2.5 Å originate from the second coordination sphere (blue), distances between carbon atoms bonded to the same carbon (e.g. ($C_3,C_5$)). The 2.5 Å maximum exhibits a shoulder towards largerdistances due to contributions from the third carbon coordination sphere (pink, distances between carbon atoms connected to a common carbon-carbon bond, e.g. ($C_3,C_{10}$)). As noted previously,[12] the third carbon coordination sphere has a bimodal distance distribution in rigid ring systems due to cis (e.g. ($C_3,C_9$)) and trans (e.g. ($C_3,C_{10}$)) conformations about the central bond. The broad tail of the PDFs towards distances beyond 4 Å results from distances in higher coordination spheres.

A difference PDF (ΔPDF), which is the difference between delay-dependent PDF and static PDF, obtained 550 fs after photoexcitation by a femtosecond pulse with a wavelength of 266 nm, when the ring-opening is already completed, is shown in **Fig. 2b**. Due to their differential nature and as demonstrated by the difference carbon-carbon distance distributions (orange in **Fig. 2b**), a distance change appears as a combination of a negative contribution to the ΔPDF at the original distance and a positive contribution at the distance, which is reached at the delay time of the ΔPDF. The ΔPDF shows the strongest features in distance regimes labeled with α, β, and ɣ in **Fig. 2b**. The α and β regions closely resemble the positions of the first and second coordination sphere in **Fig. 2a**. Since the α and β signals exhibit negative amplitudes, they correspond to the disappearance of distances of the reactant first and second coordination sphere. Such signatures are only consistent with ring-opening.[12,15] The bond dissociation as part of the ring-opening increases the ($C_3$-$C_4$) distance and, therefore, leads to a negative signature in the α region. The β region is dominated by negative signatures from increases in the ($C_1$,$C_4$) and ($C_3$,$C_5$) distances. All three of these distances increase during the ring-opening reaction towards values in the γ regime and beyond, leading to a positive signature there. The ΔPDF exhibits additional negative and positive signatures in the 4-5 Å and 5-6 Å regions which result from ring-opening induced (C,C) distance changes in the third and higher coordination spheres.

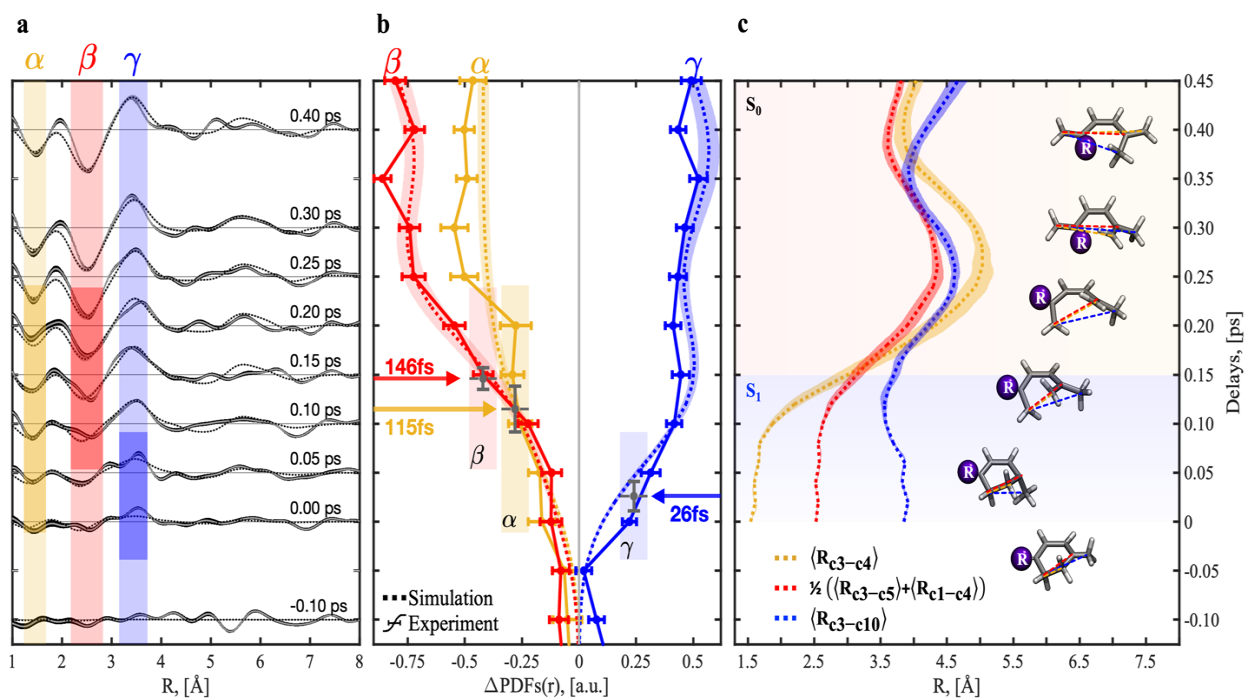

**Figure 3. Comparison of measured and simulated time-dependent difference pair distribution functions (ΔPDFs).** Panel (a) shows experimental and simulated (solid and dotted black lines) ΔPDFs at different delays in a time window of 0.5 picoseconds (ps) around time-zero. Simulations use the *ab initio* multiple spawning (AIMS) method (see **methods**). Analogous to **Fig. 2**, α, β, and γ regions are highlighted with different colors. Panel (b) shows the integrated experimental (solid) and simulated (dotted) time-dependent intensity of the three regions from panel (a) with identical color-coding. We fit error functions to the temporal onsets of the experimental signals (see **Supplementary Note 2**). The delay values of the error function centers are marked with arrows and as black dots with error bars. Additionally, the widths of the signal onsets according to the error function fits are marked by color-coded bars in panels (a) and (b). The shaded areas (simulation) and error bars (measurement) of the line plots indicate the uncertainty obtained from bootstrapping analysis (68% confidence interval). For the simulations, these error bars reflect convergence with respect to initial condition sampling. The temporal evolution of three representative carbon-carbon distances in the AIMS simulations, the $C_3$-$C_4$, $C_3$-$C_5$, and $C_3$-$C_{10}$ distances, (labeling according to **Fig. 2a**) is plotted in panel (c). Additionally, snapshots of the molecular geometry evolution based on a representative AIMS trajectory are shown with the three representative carbon-carbon distances marked.

We simulate the ring-opening dynamics of αTP using *ab initio* multiple spawning (AIMS)[23–25] in combination with α-state-averaged complete active-space self-consistent field theory (α-CASSCF)[26] for electronic structure determination and generate ΔPDFs from them (see **Fig. 2b**, **Figs. 3-5** and the **Methods**). According to our simulations, 58 % of the αTP excited state population relaxing through the conical intersection with $S_0$ undergoes electrocyclic ring-opening, whereas the remaining population returns to the reactant minimum. We find a high level of agreement between experimental and simulated ΔPDFs. **Figure 3a** shows in detail the temporal onset of the ΔPDF signal from **Fig. 2b** over several delay steps around time zero. The α, β, and ɣ regions are highlighted in **Fig. 3a**. The time-dependent evolution of the integrated signal from the three regions is plotted in **Fig. 3b**. We fit the temporal onset of the transient signal in all three regions with error functions (see **Supplementary Note 2**). The center and width of these error function fits are shown in **Fig. 3b**. Both experiment and simulation show a delayed onset of the α and β signatures with respect to the ɣ signature around time zero.

We have assigned the positive amplitude in the ɣ region of the temporal snapshot of **Fig. 2b** at 550 fs delay to an increase of the ($C_3$-$C_4$), ($C_1$,$C_4$), and ($C_3$,$C_5$) and other distances from the α and β to the ɣ regime. This assignment cannot hold for the early onset of the positive signature in ɣ at time zero since it precedes the onset of the corresponding negative α and β signatures. Thus, the signature must originate from structural dynamics prior to the ($C_3$-$C_4$) bond breaking and the structural opening of the ring.

We have observed in previous studies of similar rigid ring systems a collapse of the bimodal distribution of the third coordination sphere. This is due to the redistribution of the absorbed photon energy during non-adiabatic dynamics lowering the molecular rigidity.[12,27] The corresponding signatures in a ΔPDF are negative peaks at the positions of the two cis and trans maxima (~3 Å and ~4 Å, respectively) of the third coordination sphere and a positive peak in the gap of the third coordination sphere (3.4 Å), overlapping with the observed ɣ signature. However, the early signature in the ɣ region as observed in the present study agrees only partially with this expectation: We observe a clear positive signature at 3.4 Å and a weak negative signature at 4 Å, which is close to the noise level in the experimental data, but clearly visible in the simulations (see **Supplemental Figure S1**), but a corresponding negative signature at smaller distances around 3 Å is missing. Thus, the early onset of the ɣ signature must exclusively originate from a distance reduction of larger third coordination sphere distances in trans configuration (see above) and distances from higher coordination spheres. An exclusive reduction of third coordination sphere distances in trans-configuration can only be consistent with out-of-plane motion of the ($C_{10}$) "reporter" carbon of the methyl substituent (e.g., ($C_3,C_{10}$)). Out-of-plane motion of the isopropyl group would also reduce distances in cis-configuration (e.g., ($C_8,C_6$)) which is not supported by our data.

Our simulations give further evidence for such a motion as visualized at the example of the ($C_3,C_{10}$) distance in **Fig. 3c**. At 100 fs after photoexcitation, the methyl ring substituent shows significant out-of-plane displacement, which leads to a reduction of the ($C_3,C_{10}$) distance from 3.9 Å to 3.2 Å while the ($C_3$-$C_4$) and the ($C_3,C_5$) distances do not yet show substantial displacements. The ($C_3$-$C_4$) distance only shows considerable enlargement after 150 fs. With the ($C_3$-$C_4$) distance increase, the ($C_3,C_{10}$) pair contributes to the ΔPDF at higher distances. The contributions to the ɣ regime are taken over by the ($C_3$-$C_4$), ($C_3,C_5$), and other carbon pairs not highlighted in **Fig. 3c**. Thus, we observe in both experiment and simulation a clear temporal separation between the methyl group out-of-plane bending, which leads to the early rise of the γ signature in the ΔPDFs, and the structural opening of the ring, which leads to the delayed onset of the α and β signatures in the ΔPDFs.

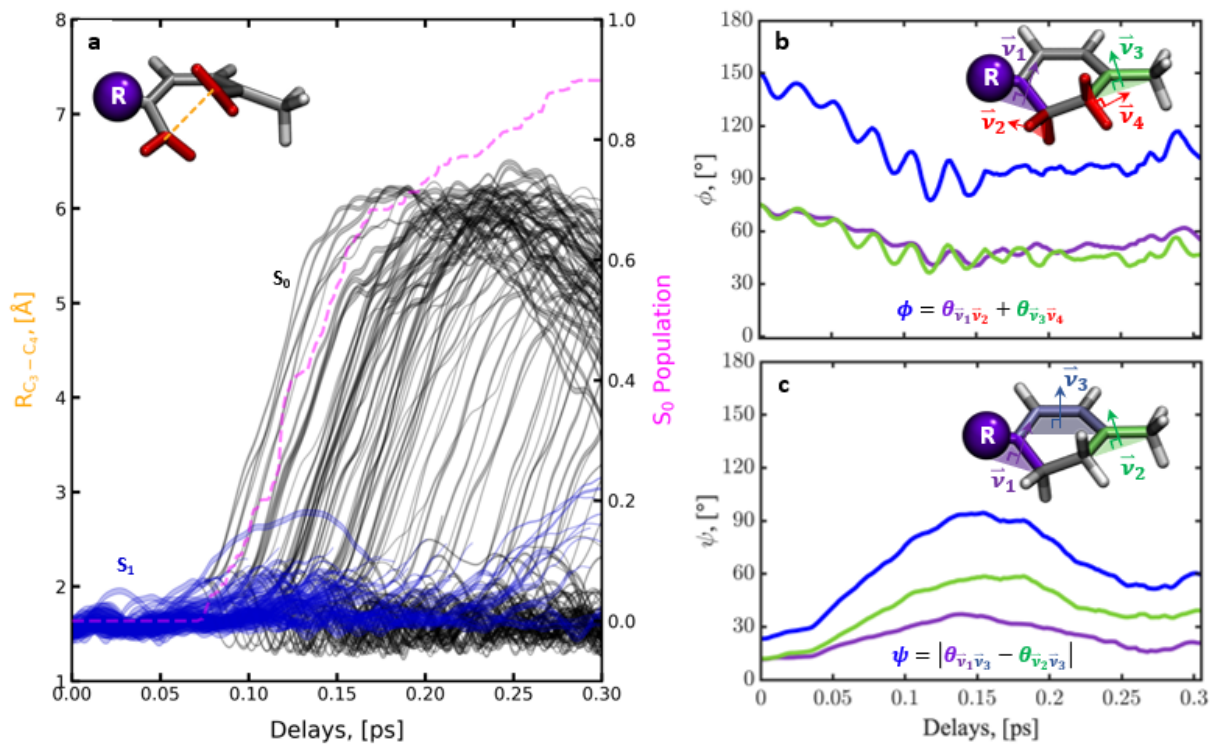

Figure 4. Temporal evolution of the simulated trajectories in several nuclear degrees of freedom. Panel a shows the evolution of the ($C_3$-$C_4$) distance in the excited state ($S_1$, blue) and the ground state ($S_0$, black). Additionally, the time-dependent population of $S_0$ is plotted (pink). Bond dissociation, i.e. ring-opening, happens directly after internal conversion from $S_1$ to $S_0$. Panel b shows the time-dependent expectation value of the projection of the simulated nuclear wavepacket evolution in $S_1$ onto the conrotatory planarization coordinate $\phi$ (blue). The coordinate is defined in the inset and represents the conrotatory addition of the angles between the plane defined by the ($C_3$) $CH_2$ group ($\vec{v_2}$) and the plane defined by the $C_1$, $C_2$, and $C_3$ carbons ($\vec{v_1}$, purple plot), and between the planes defined by the ($C_4$) $CH_2$ group ($\vec{v_4}$) and the plane defined by the $C_4$, $C_5$, and $C_{10}$ carbons ($\vec{v_3}$, green curve), respectively. The corresponding projections onto a conrotatory deplanarization coordinate $\psi$ (blue) is plotted in panel c. The coordinate is defined in the inset and represents the conrotatory addition of the angles of the planes defined by the $C_1$, $C_2$, and $C_3$ carbons ($\vec{v_1}$, purple curve) and the $C_4$, $C_5$, and $C_{10}$ carbons ($\vec{v_2}$, green curve) with respect to a common plane defined by the $C_1$, $C_6$, $C_9$, and $C_5$ ($\vec{v_3}$).

**Figure 4a** shows a projection of the excited state (blue) and ground state (black) components of the trajectory representations of the simulated wavepacket onto the ($C_3$-$C_4$) distance. The projection clearly shows that ($C_3$-$C_4$) bond dissociation happens exclusively in the ground state and quasi-instantaneously after population transfer to the ground state through the CI (see **Fig.1**). Thus, the methyl group out-of-plane bending must take place in the excited state prior to internal conversion through the conical intersection with the ground state. Hence, it is a direct signature of the structural relaxation to the pericyclic minimum of $S_1$.

The out-of-plane bending can be directly related to conrotatory rehybridization dynamics enabling interaction between $\pi$ and $\sigma$ electrons of the molecule. Rehybridization of the ($C_3$) and ($C_4$) $CH_2$ groups from $sp^3$ to $sp^2$ hybridization must lead to a planarization around the terminal double bonds of the photoproduct (($C_3=C_1$) and ($C_4=C_5$), see **Fig. 1**), i.e. moving the ($C_3$) $CH_2$ group into a common plane with the ($C_1$), ($C_2$), and ($C_6$) carbons and the ($C_4$) $CH_2$ group into a common plane with the ($C_5$), ($C_9$), and ($C_{10}$) carbons, respectively. Such a planarization could be achieved by the conrotatory movement of the hydrogens around the respective carbons, in line with the simplified picture given by Woodward and Hoffmann.[2] However, it is strongly restricted by the presence of the still intact ($C_3$-$C_4$) $\sigma$ bond in $S_1$ (see **Supplementary Note 3**).

As an alternative, planarization with respect to the terminal double bonds of the photoproduct can be achieved by deplanarization of the methyl and isopropyl substituents with respect to the conjugated ($C_1=C_6$) and ($C_9=C_5$) double bonds of the reactant. In **Fig. 4b** and **c**, we plot the expectation value of the excited state component of the simulated wavepacket onto the corresponding degrees of freedom. **Figure 4b** depicts a projection onto the conrotatory planarization coordinate $\phi$ with respect to the terminal ($C_1=C_3$) and ($C_4=C_5$) double bonds of the photoproduct involving the methyl and isopropyl substituents. The projection of the excited state wavepacket onto the complementary conrotatory deplanarization coordinate $\psi$ with respect to the cis-butadiene-like conjugated double bond system of the reactant is plotted in **Fig. 4c**. The simulated excited state wavepacket shows substantial evolution in both the planarization and deplanarization coordinates and confirms the out-of-plane motion to be dominated by the methyl group (see additional details in **Supplementary Note 4**). Additionally, the minimum (maximum) points of the motions in **Figs. 4b** (4c) temporally coincide well with depopulation to the electronic ground state (see the pink curve in **Fig. 4a**). This finding strongly suggests a connection between the deplanarization/planarization motion, i.e., π bond alternation and $CH_2$ rehybridization, and the access to the conical intersection seam in the vicinity of the pericyclic minimum.

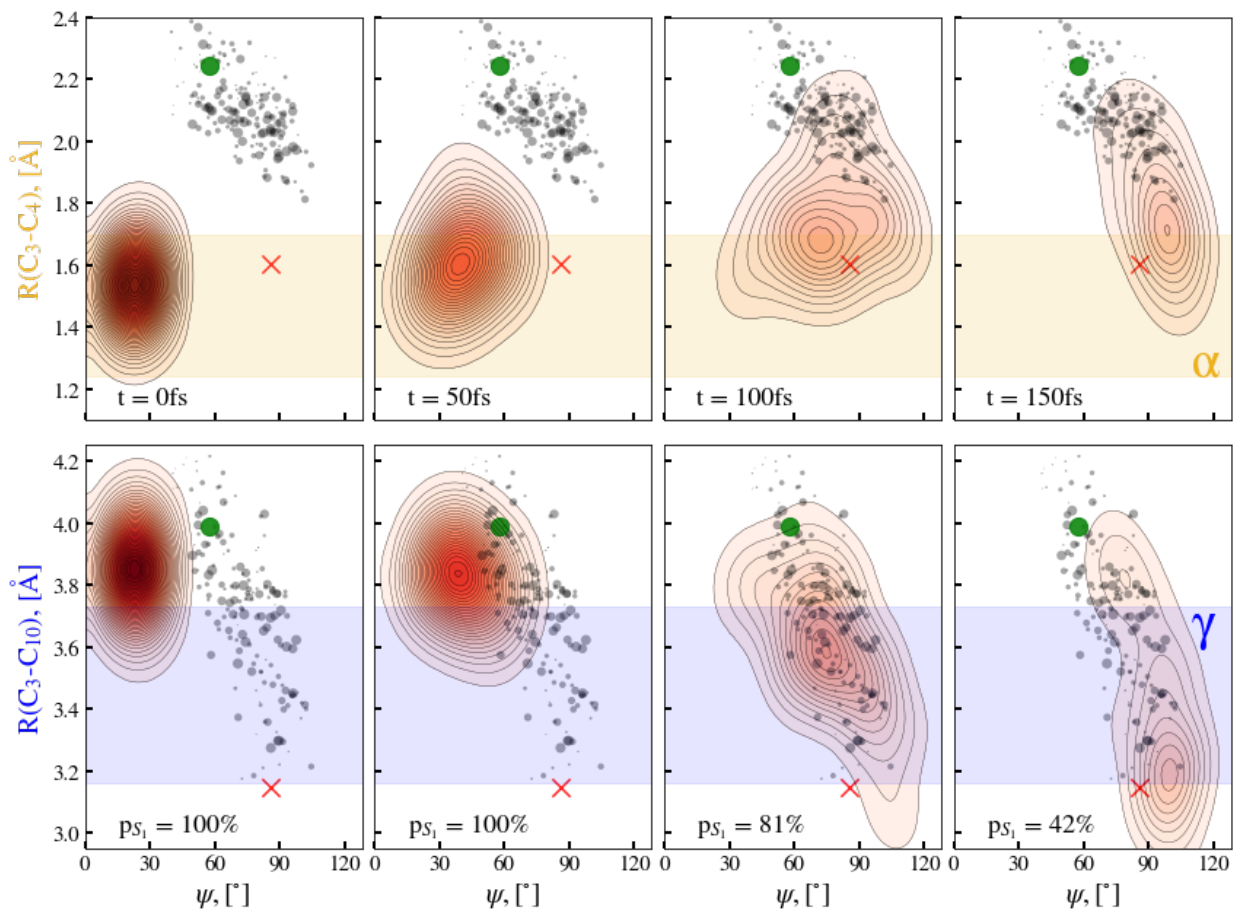

**Figure 5. Two-Dimensional projections of the simulated excited state wavepacket density.** Projections (red) onto the ($C_3$-$C_4$) distance and the conrotatory deplanarization angle $\psi$ from **Fig. 4c** for different time delays are shown in the upper row. Analogous projections onto the ($C_3$,$C_{10}$) distance and $\psi$ are depicted in the lower row. The delays for each column are written in the upper row plots, the population fraction residing in the excited state is marked in the lower row plots. For comparison, the α and γ areas of **Fig.** 3 are marked in yellow and blue. Additionally, the geometries at which population transfer to the ground state takes place are shown as grey circles with sizes proportional to the relative amount of transferred population. The minimum energy conical intersection geometry is marked as a green circle and the geometry of the pericyclic minimum as a red cross.

The upper row of **Fig. 5** shows two-dimensional projections of the simulated excited state wavepacket density (red) onto the ($C_3$-$C_4$) distance and the deplanarization angle $\psi$ from **Fig. 4c** at different delay times. Corresponding projections onto the ($C_3$-$C_{10}$) distance and ψ are plotted in the lower row of **Fig. 5**. The projection shows significant anti-correlation between the ($C_3$-$C_{10}$) distance and ψ (lower row), specifically at the onset of population transfer to the ground state (100 fs delay). In contrast, there is neither strong correlation nor anti-correlation for the ($C_3$-$C_4$) distance and ψ (upper row). Additionally, significant motion of the excited state wavepacket density from outside into the γ regime of the ΔPDFs (blue-shaded area, as defined in **Fig. 3**) can be seen in the lower-row graphs, whereas the density maximum of the excited wavepacket barely leaves the α-regime (yellow-shaded area, upper row). Thus, the early onset of the amplitude increase within the γ regime of the experimental ΔPDFs (**Fig. 3**) can be regarded as a unique and sensitive gauge for the conrotatory deplanarization in the molecule and, thus, for the rehybridization of the ($C_3$) and ($C_4$) $CH_2$ groups.

We compare the wavepacket evolution in **Fig. 5** with two significant points identified in our theoretical investigations of the $S_1$ potential energy surface, the pericyclic minimum geometry (red cross in **Fig. 5** and molecular geometry in **Fig. 1**) and the minimum energy conical intersection geometry (MECI, green circle in **Fig. 5** and molecular geometry in **Fig. 1**). The MECI is separated from the pericyclic minimum by a small barrier. Both geometries show significant out-of-plane bending. The wavepacket motion in both projections of **Fig. 5** is clearly driven by a gradient in the Franck-Condon region of $S_1$ (0 fs delay in **Fig. 5**) towards the pericyclic minimum. In the vicinity of the pericyclic minimum, it encounters a region of the $S_1$ potential energy surface with strong nonadiabatic coupling leading it to undergo internal conversion to $S_0$ (grey circles in **Fig. 5**). Thus, **Fig. 5** nicely demonstrates that the conical intersection seam is the origin of the strong non-adiabatic coupling which drives internal conversion and subsequent ring-opening, but that nonadiabatic transitions do not necessarily happen exactly at the conical intersection seam or the MECI.

In conclusion, by the combination of ultrafast electron diffraction and AIMS simulations we provide a detailed molecular picture of the rehybridization dynamics to the photochemical "transition state" of an electrocyclic reaction, the pericyclic minimum. The pericyclic minimum represents different levels of progress for the multiple concerted processes involved in the reaction, σ bond dissociation, π-bond alternation, and rehybridization. We observe a significant level of rehybridization and π-bond alternation happening during relaxation in $S_1$ towards the pericyclic minimum. However, the structural motion during the relaxation can only be explained by rehybridization in the presence of an intact σ bond. Thus, the pericyclic minimum represents an early stage of the reaction with respect to σ bond dissociation and a significantly later stage with respect to the other processes. Our results provide a new perspective on the origins of the stereospecificity of electrocyclic reactions: The stereoconfiguration of the photoproduct is rather preserved by excited state rehybridization dynamics in the presence of the σ bond locking the double bond structure of the photoproduct in place rather than by a conrotatory motion of the $CH_2$ groups during the σ bond dissociation.

**Acknowledgements**

We thank Markus Gühr, Dennis Mayer, and Stephen Weathersby for their support of the experiment and helpful discussions. This work was supported by the AMOS program within the U.S. Department of Energy, Office of Science, Basic Energy Sciences, Chemical Sciences, Geosciences, and Biosciences Division. MeV-UED is operated as part of the Linac Coherent Light Source at the SLAC National Accelerator Laboratory, supported in part by the U.S. Department of Energy (DOE) Office of Science, Office of Basic Energy Sciences, SUF Division Accelerator and Detector R&D program, the LCLS Facility, and SLAC under contract Nos. DE-AC02-05CH11231 and DE-AC02-76SF00515. Lawrence Livermore National Laboratory is operated by Lawrence Livermore National Security, LLC, for the U.S. Department of Energy, National Nuclear Security Administration, under Contract DE-AC52-07NA27344. J.P.F.N. and M.C. were supported by the US Department of Energy Office of Science, Basic Energy Sciences under award no. DE-SC0014170.## References

1. Havinga, E. & Schlatmann, J. L. M. A. Remarks on the specificities of the photochemical and thermal transformations in the vitamin D field. *Tetrahedron* **16**, 146–152 (1961).

2. Woodward, R. B. & Hoffmann, R. The Conservation of Orbital Symmetry. *Angew. Chem. Int. Ed.* **8**, 781 (1969).

3. Manolopoulos, D. E. *et al.* The Transition State of the F + H2 Reaction. *Science* **262**, 1852–1855 (1993).

4. Pedersen, S., Herek, J. L. & Zewail, A. H. The Validity of the 'Diradical' Hypothesis: Direct Femtoscond Studies of the Transition-State Structures. *Science* **266**, 1359–1364 (1994).

5. Wenthold, P. G., Hrovat, D. A., Borden, W. T. & Lineberger, W. C. Transition-State Spectroscopy of Cyclooctatetraene. *Science* **272**, 1456–1459 (1996).

6. Weichman, M. L. *et al.* Feshbach resonances in the exit channel of the F + CH3OH → HF + CH3O reaction observed using transition-state spectroscopy. *Nat. Chem.* **9**, 950–955 (2017).

7. Garavelli, M. *et al.* Reaction Path of a sub-200 fs Photochemical Electrocyclic Reaction. *J. Phys. Chem. A* **105**, 4458–4469 (2001).

8. Deb, S. & Weber, P. M. The Ultrafast Pathway of Photon-Induced Electrocyclic Ring-Opening Reactions: The Case of 1,3-Cyclohexadiene. *Annu. Rev. Phys. Chem.* **62**, 19–39 (2011).

9. Lugt, W. T. A. M. van der & Oosterhoff, L. J. Quantum-chemical interpretation of photo-induced electrocyclic reactions. *Chem. Commun. Lond.* 1235–1236 (1968) doi:10.1039/C19680001235.

10. Van der Lugt, W. Th. A. M. & Oosterhoff, L. J. Symmetry control and photoinduced reactions. *J. Am. Chem. Soc.* **91**, 6042–6049 (1969).

11. Michl, J. Model calculations of photochemical reactivity. *Pure Appl. Chem.* **41**, 507–534 (1975).

12. Champenois, E. G. *et al.* Conformer-specific photochemistry imaged in real space and time. *Science* **374**, 178–182 (2021).

13. Attar, A. R. *et al.* Femtosecond x-ray spectroscopy of an electrocyclic ring-opening reaction. *Science* **356**, 54–59 (2017).

14. Yang, J. *et al.* Imaging CF3I conical intersection and photodissociation dynamics with ultrafast electron diffraction. *Science* **361**, 64–67 (2018).

15. Wolf, T. J. A. *et al.* The photochemical ring-opening of 1,3-cyclohexadiene imaged by ultrafast electron diffraction. *Nat. Chem.* **11**, 504–509 (2019).

16. Centurion, M., Wolf, T. J. A. & Yang, J. Ultrafast Imaging of Molecules with Electron Diffraction. *Annu. Rev. Phys. Chem.* **73**, 21 (2022).

17. Arruda, B. C., Peng, J., Smith, B., Spears, K. G. & Sension, R. J. Photochemical Ring-Opening and Ground State Relaxation in α-Terpinene with Comparison to Provitamin D3. *J. Phys. Chem. B* **117**, 4696–4704 (2013).

18. Arruda, B. C., Smith, B., Spears, K. G. & Sension, R. J. Ultrafast ring-opening reactions: a comparison of α-terpinene, α-phellandrene, and 7-dehydrocholesterol with 1,3-cyclohexadiene. *Faraday Discuss.* **163**, 159–171 (2013).

19. Arruda, B. C. & Sension, R. J. Ultrafast polyene dynamics: the ring opening of 1,3-cyclohexadiene derivatives. *Phys. Chem. Chem. Phys.* **16**, 4439–4455 (2014).

20. Gao, Y., Pemberton, C. C., Zhang, Y. & Weber, P. M. On the ultrafast photo-induced dynamics of α-terpinene. *J. Chem. Phys.* **144**, 194303 (2016).

21. Krause, P., Matsika, S., Kotur, M. & Weinacht, T. The influence of excited state topology on wavepacket delocalization in the relaxation of photoexcited polyatomic molecules. *J. Chem. Phys.* **137**, 22A537 (2012).

22. Minitti, M. P. *et al.* Imaging Molecular Motion: Femtosecond X-Ray Scattering of an Electrocyclic Chemical Reaction. *Phys. Rev. Lett.* **114**, 255501 (2015).

23. Ben-Nun, M. & Martínez, T. J. *Adv Chem Phys* **121**, 439–512 (2002).

24. Ben-Nun, M., Quenneville, J. & Martínez, T. J. Ab Initio Multiple Spawning: Photochemistry from First Principles Quantum Molecular Dynamics. *J. Phys. Chem. A* **104**, 5161–5175 (2000).

25. Ben-Nun, M. & Martínez, T. J. Nonadiabatic molecular dynamics: Validation of the multiple spawning method for a multidimensional problem. *J. Chem. Phys.* **108**, 7244–7257 (1998).

26. Snyder, J. W., Parrish, R. M. & Martínez, T. J. α-CASSCF: An Efficient, Empirical Correction for SA-CASSCF To Closely Approximate MS-CASPT2 Potential Energy Surfaces. *J. Phys. Chem. Lett.* **8**, 2432–2437 (2017).

27. Nunes, J. P. F. *et al.* Observing the photo-induced structural dynamics of proton transfer in o-nitrophenol by Ultrafast Electron Diffraction. Preprint at https://doi.org/10.21203/rs.3.rs-1837872/v1 (2022).

# SUPPLEMENTARY INFORMATION FOR

# REHYBRIDIZATION DYNAMICS INTO THE PERICYCLIC MINIMUM OF AN ELECTROCYCLIC REACTION IMAGED IN REAL-TIME

Y. Liu,[1,2,†] D. M. Sanchez,[1,3,†,‡], M. R. Ware,[1] E. G. Champenois,[1] J. Yang,[1,4,5] J. P. F. Nunes,[6,7] A. Attar,[4] M. Centurion,[6] J. P. Cryan,[1] R. Forbes,[4] K. Hegazy,[1] M. C. Hoffmann,[4] F. Ji,[4] M.-F. Lin,[4] D. Luo,[4] S. K. Saha,[6] X. Shen,[4] X. J. Wang,[4,] T. J. Martínez,[1,3,*] & T. J. A. Wolf[1,*]

[1]Stanford PULSE Institute, SLAC National Accelerator Laboratory, Menlo Park, USA.

[2]Department of Physics and Astronomy, Stony Brook University, Stony Brook, NY, USA.

[3]Department of Chemistry, Stanford University, Stanford, USA.

[4]SLAC National Accelerator Laboratory, Menlo Park, USA.

[5]Center of Basic Molecular Science, Department of Chemistry, Tsinghua University, Beijing, China.

[6]Department of Physics and Astronomy, University of Nebraska-Lincoln, Lincoln, USA.

[7]Diamond Light Source, Harwell Science Campus, Didcot, UK

*Corresponding authors. Emails: thomas.wolf@slac.stanford.edu, toddjmartinez@gmail.com

† These authors contributed equally to this work.

‡Present address: Design Physics Division, Lawrence Livermore National Laboratory, Livermore, USA.

**TABLE OF CONTENTS**

**METHODS**

**Experimental Setup**: The experimental apparatus is described in detail elsewhere.[1,2] In short, we use the 800 nm, 50 fs output of a Ti:Sapphire laser system and separate two beam paths. Pulses in both beam paths are frequency tripled. The pulses of the probe beam path are directed onto the photocathode of an RF gun and eject an ultrashort pulse containing ~$10^4$ electrons. 4.2 MeV electrons are generated using a S-band photocathode radio frequency (RF) gun[2] and focused to a spot size of 200 μm FWHM in the interaction region of a gas phase experimental chamber. The pump pulses (4-6 μJ) are focused into the experimental chamber to a diameter of 240 μm FWHM and overlapped with the electron pulses at a 1° angle through a holey mirror in the interaction region of a gas phase experimental chamber. The experimental response function including effects of the optical and electron pulse length as well as relative arrival time jitter is estimated to be 150 fs.[3,4] α-Terpinene (αTP, purity >95%) is purchased from Sigma-Aldrich and used without further purification. We employ a static-filled 3 mm flow cell (550 μm orifices, sample at room temperature) in combination with a repetition rate of 360 Hz. Diffracted electrons are detected by a combination of a phosphor screen and an EMCCD camera. Based on the relative static and dynamic signal levels, we estimate that we excite about 1.56 % of the molecules (for details see **Ultrafast Electron Diffraction Signal Processing**) Time-dependent diffraction is measured at a series of time delay points between -2 ps and +2 ps in each scan. The separation between time delay points is 50 fs, except for the earliest and latest delay points, where it is considerably larger. At each time delay point, we integrate the diffraction signal for 10 seconds (3600 shots), and the full data set includes 145 such scans at each delay. The sequence of delay steps is randomized for every scan to avoid systematic errors. The processing of the obtained raw diffraction patterns is described in detail in the **Ultrafast Electron Diffraction Signal Processing** section.

**Excited-State Dynamics Simulations:** Ab initio multiple spawning (AIMS) simulations interfaced with GPU-accelerated α-Complete Active Space Self-Consistent Field Theory (α-CASSCF)[5–7] is used to model the photochemical ring-opening dynamics of αTP in isolation. In its ring-closed form, αTP exhibits three relatively stable and optically indistinguishable rotamers which differ by the orientation of the isopropyl group with respect to the ring plane. For this reason, all are included (with equal weight) in the AIMS simulations using identical methods. For more details about the different structural signatures of the rotamers see **Supplementary Note 5**. Our active space consists of six electrons in four orbitals (1σ*, 2π, and 1π*) and the two lowest energy singlet states (referred to as $S_0$ and $S_1$ for the ground and 1st excited state in the adiabatic representation, respectively), within the 6-31G* basis set using an α value of 0.82, i.e. α(0.82)-SA2-CAS(6,4)SCF/6-31G*. Based on our previous work with 1,3-cyclohexadiene (CHD) and α-phellandrene (αPH), it is expected that this level of theory is well-suited for describing the photoinduced ring-opening dynamics of αTP.[4,8] Similar to CHD and αPH, there exist both closed- and ring-open $S_0/S_1$ conical intersections (CI) in αTP, which correspond to geometries where the CHD moiety puckers out of plane (OOP) or the $C_3$-$C_4$ bond (**Figs. 2a** and **S2**) elongates to approximately 2.2 Å, respectively. **Fig. S2** shows the "critical points" (i.e. stationary points) upon evolving away from the Frank Condon point on $S_1$ along the open and closed photochemical pathways for the three rotamers of αTP. As in CHD and αPH, the ring-open CI is lower in energy than the closed-ring CI, offering some insight into a branching ratio that heavily favors the ring-opening pathway (see discussion on **αTP Photoinduced Ring-Opening Dynamics**). In addition, we observe a minimum on $S_1$ near the ring-open CI, where the methyl group moves OOP of the CHD moiety and the $C_3$-$C_4$ bond elongates to 1.6 Å. This is consistent with the pericyclic minimum in photoinduced electrocyclic reactions; an intermediate on $S_1$ that leads to ring-opening dynamics via nonradiative decay through the ring-open CI. All α-CASSCF electronic structure calculations are performed with the TeraChem electronic structure package.[9–11]

**Fig. S3a** shows an ultraviolet (UV) electronic absorption spectrum generated from 300 geometries (100 per αTP rotamer) sampled from a harmonic Wigner distribution corresponding to the rotamer's PBE0/6-31G* ground state optimized structure. Single point energy calculations are performed at the α(0.82)-SA2-CAS(6,4)SCF/6-31G* level of theory and the resulting $S_0 \rightarrow S_1$ excitation energies homogeneously broadened using an oscillator-strength scaled Gaussian with a full width half maximum (FWHM) of 0.2 eV. The active-space molecular orbitals (MO) at the $S_0$ minima are shown in **Fig. S3b** for all rotamers, which are nearly identical to those used in our CHD and αPH work. A total of 60 initial conditions (20 sets of positions and momenta for each rotamer) are selected to initiate the AIMS dynamics. These initial conditions (ICs) are selected under the constraint that their $S_0 \rightarrow S_1$ transition energy was within 0.3 eV of the pump pulse (4.65 eV) used in the experiment after applying a 0.4 eV red-shift to the theoretical spectrum to align the theoretical and experimental absorption maxima. These initial conditions are then placed on the $S_1$ surface and propagated with AIMS.

AIMS expands the full molecular wavefunction into a time-dependent basis of multi-dimensional frozen-width Gaussian functions that evolve along adiabatic PESs according to the time-dependent molecular Schrodinger equation.

$$\chi_I(R,t) = \sum_{k=1}^{N_I(t)} c_k^I(t) \chi_k^I \left( R; \underline{R}_k^I(t), \underline{P}_k^I(t), \underline{\gamma}_k^I(t), \alpha_k^I \right) \tag{1}$$

where $N_I(t)$ represents the total number of trajectory basis functions (TBFs) on electronic state I, $c_k^I(t)$ is the time-dependent complex coefficient of the $k$th TBF, $\alpha_k^I$ is the frozen TBF width, and $\chi_k^I(...)$ is a multidimensional frozen Gaussian that is expressed as a product of one-dimensional Gaussian functions corresponding to the 3N nuclear degrees of freedom. The reader is referred elsewhere for a more detailed description of performing and analyzing excited-state dynamics simulations within the AIMS framework.[4,8,12–14] The first two singlet states ($S_0$ and $S_1$) are included in the AIMS dynamics. All required electronic structure quantities (energies, gradients, and nonadiabatic couplings) are calculated on-the-fly with α-SA2-CAS(6,4)SCF/6-31G*. An adaptive timestep of 0.48 fs (20 au) (reduced to 0.12 fs (5 au) in regions with large nonadiabatic coupling) is used to propagate the centers of the trajectory basis functions (TBFs). A coupling threshold of 0.01 au (scalar product of nonadiabatic coupling and velocity vectors) demarcates spawning events generating new TBFs on different electronic states. Population transfer between TBFs is described by solving the time-dependent Schrödinger equation in the time-evolving TBF basis set.

We simulate the ultrafast dynamics for the first 1 ps of all three rotamers of αTP by: 1) using AIMS to propagate the initial wavepacket for the first 500 fs or until all population has returned to the ground state, 2) stopping TBFs on the ground state when they are decoupled from other TBFs (off-diagonal elements of the Hamiltonian become small), and 3) adiabatically continuing these stopped TBFs using the positions and momenta from the last frame in AIMS as initial conditions for adiabatic molecular dynamics with unrestricted DFT using the Perdew-Burke-Ernzerof hybrid exchange-correlation functional,[15] i.e., uPBE0/6- 31G*. A total of 234 TBFs are propagated, with 174 of these being adiabatically continued on the ground state with DFT. The simulation of time-dependent $\Delta PDF(r,t)$ is described in detail in the section **Simulated Diffraction Signals**.

# ULTRAFAST ELECTRON DIFFRACTION SIGNAL PROCESSING

**Diffraction Signal Treatment:** After the experimental data acquisition, each measured diffraction pattern undergoes a quality evaluation and control routine which consists of several processes including, 1) signal baseline subtraction, 2) rejection of images with low signal intensity, 3) removal of "hot" pixels, and 4) median filtering and normalization. More details of the analysis procedure of the signal treatment can be found in the supplementary materials of Ref.[3,4,8,16] After the evaluation and control process, the diffraction centers of each individual image are located. The center of each image is determined by a least square fitting algorithm on the diffraction pattern. To account for the changes in the total electron number in each pulse, each diffraction pattern (or image) is normalized to the total signal in the ranges of $2 < s < 8$ Å$^{-1}$.

**Diffraction Percentage Difference:** We optimize our analysis of the experimental data by evaluating diffraction percentage difference signals. The 1-dimensional scattering intensity as a function of s, $I_{exp}(s)$, is obtained by azimuthally averaging the 2-dimensional diffraction pattern using the determined centers. To highlight the time-dependent changes in the data, we calculate the percentage difference signal according to the following equation,

$$\%\Delta \frac{I}{I_{exp}}(s,t) = \frac{I_{exp}(s,t) - I_{exp}(s,t<0)}{I_{exp}(s,t<0)} \times 100 \qquad (2)$$

where $I_{exp}(s,t)$ is the total scattering intensity for each pump-probe delay time, t. The reference signal $I_{exp}(s, t < 0)$ is obtained by averaging the diffraction signal measured at delays -2 ps < t < -500 fs, which corresponds to the static, unpumped scattering signal. **Fig. S4** shows a false-color plot of the experimental $\%\Delta \frac{I}{I_{exp}}(s,t)$ signal. **Fig. S5** panels **a** and **b** show several representative delay-slices of the measured signals (curves) and uncertainty (shaded areas) at specific pump-probe delays before and after a further background removal process, respectively. In this process, a low-order (up to 2nd order) polynomial is fitted over the whole s range.[16,17] This procedure has little effect on the low s region, but significantly helps to reduce noise and systematic offsets at high s (see **Fig. S5b).** We applied a standard bootstrapping analysis to estimate the statistical uncertainty of the measurement. In total, the data set contains 145 runs, and a single run contains diffraction patterns at each individual delay. We made use of these 145 runs to create bootstrapped datasets. In each bootstrapped dataset, 145 diffraction patterns are randomly selected per delay step and this process is repeated by 150 times, generating 150 bootstrapped datasets. Each dataset is analyzed along the same protocols such that a mean and standard deviation can be evaluated for the variables with interests. **Fig S5** panel **b** shows the error bar of the measurement which is one standard deviation of the $\%\left(\frac{\Delta I}{I}\right)_{exp}$ from the bootstrapping analysis. The amplitude of the error bar is roughly the same order of the noise level before time-zero, which is significantly smaller than the signal level in the positive delays. **Fig. S6** reflects the uncertainty development as a function of the number of bootstrapped datasets used in the analysis. The uncertainty quickly develops in the first 50 bootstrapped datasets, with larger amplitude changes in both standard deviation $\sigma_{(\%\Delta I/I)}$ and the standard deviation of the mean saturation. However, both start saturating after running 150 bootstrapped datasets.

**Excitation Ratio and Time-Zero Location:** We estimate the experimental excitation ratio by comparing experimental and simulated diffraction signals at long delay times. **Fig. S7** shows the measured $\%\left(\frac{\Delta I}{I}\right)_{exp}(s,t)$ versus the simulated diffraction percentage difference, $\%\left(\frac{\Delta I}{I}\right)_{theory}(s,t)$, averaged between 0.8 and 1.2 ps. By multiplying a factor of 0.0156, the experimental data and simulation match well. Thus, we use this value ($\gamma$ = 1.56 %) as an estimation of the excitation ratio in the experiment.

During the experiment, we obtain time zero with the help of ultrafast heating heating from a solid-state bismuth target. However, other than in time-resolved electronic spectroscopy, the onset of such a heating signal never coincides precisely with time zero, since the nuclei have to move considerably to show a transient effect. Therefore, the experimental time zero must be corrected for a quantitative comparison with simulations. This can be achieved with the help of the simulations themselves if the agreement between experiment and simulation is sufficient. Thus, the exact time zero of the experimental data is determined by comparing the the time dependence of the integrated signal in a window between 2.8 and 3.5 Å$^{-1}$ of $\%\left(\frac{\Delta I}{I}\right)_{theory}(s,t)$, broadened with a 150 fs FWHM Gaussian to account for the limited time resolution of the experiment,  with $\%\left(\frac{\Delta I}{I}\right)_{exp}(s,t)$ integrated in the same window. We use a least-square fit to overlap the resulting signals. The fit results are shown in **Fig. S8**. The best fit yields a shift of about 105 fs.

**Pair Distribution Functions:** After the diffraction pattern analysis, we calculate the time-dependent atomic pair distribution functions (PDFs) following standard procedures that were developed by Zewail and co-workers and we made use of PDFs to extract the structural information.[17–19] Instead of using the percentage difference in **Eq. 2**, we utilized the modified scattering intensity, $sM(s;t)$, to generate the map of the $PDF(r,t)$. We calculate the $\Delta sM(s,t)$ by following the equations below,

$$\Delta sM_{exp}(s,t) = \frac{I_{exp}(s,t) - I_{exp}(s,t<0)}{I_{ato}(s)} s \qquad (3)$$

$$I_{ato}(s) = \sum_{i=1}^{N} f_i^*(s) f_i(s) \qquad (4)$$

where $\Delta sM_{exp}(s,t)$ is the delay dependent modified scattering intensity, $I_{ato}(s)$ the atomic scattering, $f_i(s)$ is the scattering amplitude of the $i^{th}$ atom calculated using the ELSEPA program.[20] The $\Delta PDF$ of the measurement is then obtained by applying a sine transform of the $\Delta sM_{exp}(s,t)$ following the equation below:

$$\Delta PDF_{exp}(r;t) = \int_0^{s_{max}} \Delta sM_{exp}(s,t)\sin(sr)\, e^{-\kappa s^2} ds \tag{5}$$

where $r$ represents the pair distance in real space, $\kappa$ is the damping constant and $s_{max}$ the magnitude of the largest scattering vector that the detector can accommodate. To minimize noise and artifacts at larger $s$ values, **Eq. 5** includes a Gaussian damping function $e^{-\kappa s^2}$ which smoothly reduces the intensity of the measured signal towards zero at the larger $s$ region. Here we use $\kappa = 0.03$, which corresponds a Gaussian function with half-width-half-maximum (HWHM) $\approx 7$ Å$^{-1}$. **Fig. S1** showcases the experimental and simulated ΔPDFs. The integration ranges for the signatures in **Fig. 3b** of the main text are shown in **Fig. S9**.

**Low S Diffraction Signal Treatment:** A general issue in the real space transformation of diffraction signals is the missing signal in the low s region which is due to the hole in the center of the phosphor screen detector. Extrapolation of the signal to $s = 0$ Å$^{-1}$ is required to avoid artifacts in the $\Delta PDF_{exp}(r,t)$. We fill in the missing experimental signal with the simulated signal following the equation below,

$$\Delta sM_{exp}(s < s_{hole}, t) = \beta(t)\Delta sM_{simu}(s \leq s_{hole}, t = 1\,ps) \tag{6}$$

$$\beta(t) = \gamma \left(1 + \left(\frac{t - t_0}{\tau}\right)\right) \tag{7}$$

**Eq. 6** consists of two terms, in which $\Delta sM_{simu}(s \leq s_{hole}; t = 1\,ps)$ is the average of the simulated $\Delta sM$ in a window between 0.98 and 1 ps in the relevant low $s$ range. The second term, $\beta(t)$, sets the time-dependent amplitude of the simulation, which is constrained to follow a simple error function with the onset time $t_0$ and width $\tau$ found by curve fitting to $\Delta sM_{exp}$ at $s > s_{hole}$. By multiplying the excitation ratio, $\gamma$, the second term fulfills the amplitude matching of the small s signal. **Fig. S1a** showcases the measured ΔPDFs after the treatment of the low S signal discussed above. **Fig. S10** shows an example of the influence from the missing data in $s < 0.8 \text{ Å}^{-1}$. Panels **a** and **b** show the $\Delta sM(s)$ and corresponding $\Delta PDF(r)$ from the simulation with or without the low $s$ region substituted by simply zeros, respectively. From panel **b**, the main artifacts induced by the substitution of zeros in the low $s$ region are a global tilted offset across the whole pair distance range in the real space, with a smooth positive contribution to pair distance < 3 Å and a smooth negative contribution > 4 Å. Panels **c** and **d** reflect the measured and simulated time-dependent $\Delta PDF(r,t)$ after the substitution of the low $s$ signal with zeros. Compared with the signals in **Fig. 2** of the main text and **Fig. S1a**, it is clear that the artifact below 2 Å from the experiment (panel **c** in **Fig. S10**) is much stronger, and the additional positive signal in the simulation (panel **d** in **Fig S10**) which is not shown in **Fig 2**.

**Simulated Diffraction Signals:** In the simulated diffraction signal, the modified difference diffraction signal defined above, $\Delta sM(s,t)$, is generated from the AIMS/DFT trajectories using the independent atomic model (IAM) and converted to $\Delta PDF(r,t)$ using identical code and procedures as for the experimental data. The total diffraction signal, $I_{Mol}(s,t)$, is computed as an average over all 60 ICs (174 TBFs), where the diffraction signal for a specific IC is approximated as an incoherent sum over weighted diffraction signals from individual TBFs:

$$I_{Mol}(s,t) = \frac{1}{N_{IC}} \sum_{M=1}^{N_{IC}} \sum_{k}^{N_{TBF}^M(t)} n_k^M(t) I_{Mol}^{k,M}(s,t) \qquad (8)$$

where $N_{IC}$ is the number of ICs, $N_{TBF}^M(t)$ is the number of TBFs at time $t$ for the $M^{\text{th}}$ IC, $n_k^M(t)$ and $I_{Mol}^{k,M}(s,t)$ are the weight and diffraction signal for the $k$th TBF of the $M^{\text{th}}$ IC at time $t$, respectively. The expression for $I_{Mol}^{k,M}(s,t)$ is identical to that used in the experimental diffraction signal, augmented with a Gaussian factor to account for the finite width of the TBFs:[21]

$$I_{Mol}^{k,M}(s,t) = \sum_i \sum_{j \neq i} |f_i(s)| \cdot |f_j(s)| \cos(\eta_i - \eta_j) \frac{\sin(s \cdot R_{ij}(t))}{s \cdot R_{ij}(t)} e^{-(\alpha_i^2 + \alpha_j^2)s^2} \qquad (9)$$

where $\alpha_i$ and $\alpha_j$ represent the finite widths for the atoms used in the TBFs.[21] These widths are taken to be element specific and are 0.112 Å/0.249 Å for carbon/hydrogen. The atomic form factors, $f_i(s)$, are calculated from ELSEPA. $R_{ij}$ is the interatomic distance between the $i^{th}$ and $j^{th}$ atoms taken from the centroids of the AIMS/DFT TBFs. The weight $n_k^M(t)$ is evaluated according to the bra-ket averaged Taylor expansion (BAT) method:[22]

$$n_k^M(t) = \frac{1}{2} \sum_l^{N_I^M(t)} [c_k^*(t) S_{kl} c_l(t) + c_l^*(t) S_{lk} c_k(t)] \qquad (10)$$

The complex amplitudes are time-independent during the DFT adiabatic dynamics and are held constant at the value from the last frame of their corresponding AIMS trajectory. This is valid because the ground state TBFs are effectively uncoupled from all other TBFs.

The theoretical $\Delta sM(s,t)$ analog to the experimental diffraction signal is computed by subtracting the static diffraction of αTP from all 2 fs time-bins. In the case of the simulations, this is the diffraction of the initial conditions, $\Delta sM(s, t = 0)$. The $\Delta sM(s,t)$ is convolved with a 150 fs full width at half maximum temporal Gaussian to match the experimental instrument response function. We use the same scripts as for the experimental data to generate $\Delta PDFs$. The error bars in **Fig. 3** of the main text represent the standard deviation of the ΔPDF values as evaluated by bootstrap sampling from the initial conditions. They are a measure for the level of convergence of the simulation for the employed number of ICs.

### αTP PHOTOINDUCED RING-OPENING DYNAMICS

**Excited State Dynamics on S$_1$:** The photoinduced ring-opening of αTP follows very closely to its parent molecule, CHD. Like CHD, the S$_1$ and S$_2$ adiabats in the FC region of αTP exhibit the diabatic character of a single and double electron excitation from its highest occupied molecular orbital (HOMO) to its lowest unoccupied molecular orbital (LUMO), respectively (see CI coefficients in min $S_0$ of **Table S1**). The $S_1$ adiabat changes character as the wavepacket evolves away from the FC region (See CI coefficients in $S_0/S_1$ MECI of **Table S1**), implying there is a CI between $S_2$ and $S_1$. As we have shown in our previous work, the $S_2/S_1$ CI is considerably sloped and will be almost avoided entirely by the wavepacket as αTP evolves towards the $S_0/S_1$ CI. Therefore, we are confident that αTP's relaxation mechanism can be described in its entirety within two adiabatic states. For this reason, all three rotamers of αTP were placed on $S_1$ to simulate its photoinduced ring-opening and the subsequent ground state isomerization dynamics in isolation for the first 1 ps.

All three rotamers of αTP show essentially identical photoinduced ring-opening dynamics and resemble CHD quite closely. **Fig. S11a** shows the population dynamics of all three rotamers as TBFs are spawned and population is transferred from $S_1$ to $S_0$. The decay constant for the photoinduced ring-opening process is 168 +/- 22 fs when averaged over all 60 ICs, which is slightly slower than CHD (139 +/- 25 fs) and faster than αPH (~456 +/- 115 fs (axial) and 285 +/- 71 fs (equatorial)). The orientation of the isopropyl group does not seem to influence the decay time (**Fig. S11a**). Furthermore, **Fig. S11b** shows that population transfer events mainly take place in the first 200 fs of dynamics when the $C_3$-$C_4$ distance is elongated past 1.8 Å as the wavepacket traverses through the $S_0/S_1$ CI. **Fig. S11c** shows that the majority of spawning geometries are quite similar in structure and energy (within 0.1 eV) to the ring-open MECI. Therefore, we treat all rotamers on an equal footing when computing observables from the AIMS dynamics.

αTP strictly relaxes through the ring-opening nonradiative relaxation pathway via a conrotatory fashion. **Table S2** shows the branching ratio from all 60 ICs between the closed- and open CIs, with effectively 100% of the wavepacket reaching the $S_0/S_1$ CI. This allows us to directly follow the conrotatory and disrotatory ring-opening motion for the entire wavepacket. **Fig. S12** shows the projection of the wavepacket population onto conrotatory and disrotatory angles (see **Fig. S12** inset) for the first 80 fs of the excited state dynamics. Increasing conrotatory/disrotatory angle means the ring opens in a conrotatory/disrotatory fashion. After excitation, the conrotatory angle increases dramatically, while the disrotatory angle remains relatively small. When the wavepacket reaches the $S_0/S_1$ CI, it can either return to the photoreactant αTP or go on to form the photoproduct cZc IPMHT. As is the case with CHD, the $S_0/S_1$ CI is relatively peaked for all rotamers of αTP (**Fig. S13**). **Tables S2** shows that approximately 42 +/- 4 % of the wavepacket results in the formation of αTP while 58 +/- 4 % forms cZc-IPMHT. This relatively split branching ratio can be attributed to the peak-like of the $S_0/S_1$ CI. Within the bootstrapped error bars, we found no single rotamer to be remarkably different.

**Ground State Dynamics on $S_0$:** The natural evolution of the wavepacket dynamics on the ground state can be followed by binning geometries along ground-state TBFs into one of the αTP photoproducts (**Fig. 1**). Following previous studies,[4,8] snapshots taken every 50 fs along all 174-ground state TBFs were binned into one of the isomers based on dihedral angles ($\Phi_1$, $\Phi_2$, and $\Phi_3$) and atomic distance $C_3$-$C_4$ ($R_1$) (**Fig. S14**) Due to the ground state TBFs being sufficiently uncoupled from all others, **Eq. 10** can be reduced to:

$$n_k^M(t) = |c_k^M(t)|^2 \tag{11}$$

where the total population of a specific isomer $L$ at time t on the ground-state, $P_L(t)$ is computed by:

$$P_L(t) = \frac{1}{N_{IC}} \sum_{M=1}^{N_{IC}} \frac{\sum_k^{N_{TBF}^M(t)} c_{k,M}^*(t)\, c_{k,M}(t)\, \delta\left(L, I\left(\underline{R}_{k,M}(t)\right)\right)}{\sum_k^{N_{TBF}^M(t)} c_{k,M}^*(t)\, c_{k,M}(t)} \tag{12}$$

where $L$ is defined as (cZc, tZc/cZt, tZt, and αTP), $c_{k,M}(t)$ is the amplitude of the $k$th TBF and $M$th IC, $N_{IC}$ is the total number of initial conditions, and $\delta(…)$ is a Kronecker-delta function, and $I(R)$ represents the isomer classification of the geometry given by **R**.

**Figs. S15a-d** shows how the excited state population relaxes into different ground-state isomers via the ring-open CI. Following the total wavefunction from all 60 ICs (**Fig. 15a**), nearly all of the ground state population is either αTP or cZc-IPMHT in the first 150fs after photoexcitation. Like the individual rotamers in **Figs. 15b-d**, the cZc-IPMHT population increases slightly quicker when compared to αTP, but quickly decreases upon the formation of the cZt/tZc-IPMHT isomers. In all panels, the cZt/tZc-IPMHT isomers are initially short lived, as the terminal ethylene groups in the "hot" ground state mixture quickly isomerize to form tZt-IPMHT. However, we only observe a large tZt-IPMHT peak initially, with the steric clash from the inward facing isopropyl group limiting its formation at later times. Similar to CHD and αPH, a small portion of the population is effectively "trapped" in cZt/tZc-IPMHT with each subsequent transformation between cZc and tZt-IPMHT. Lastly, the αTP population gradually approaches a limit between 40-50%, where the population is unable to overcome the large ground state barrier separating αTP and cZc-IPMHT.

**SUPPLMENTARY NOTES**

**Note 1, Signatures from Hydrogens in Pair Distribution functions:** The contributions from (C,H) and (H,H) distances to the static PDF of αTP are considerable smaller than those from (C,C) distances, but not negligible (see **Fig. S16**, top). However, as we show at the example of a simulated ΔPDF based on an optimized photoproduct and a reactant geometry (**Fig. S16**, bottom), The contributions from (C,C) and (C,H) distances to transient signatures are considerably smaller than to static signatures. This effect can be explained by the fact that the hydrogens closely follow the carbons, to which they are bound.

**Note 2, Error Function Fits to Temporal Onsets of the ΔPDF Signal in regions α, β, and γ:** We characterize the temporal onset of the integrated ΔPDF Signal the α, β, and γ regions by fitting an error function

$$\Delta PDF(t) = A + B \cdot erf\left(2 \cdot \sqrt{\ln(2)}\,\frac{t-t_0}{\tau}\right)$$

to it, where A is a constant offset, B the signal amplitude, $t_0$ the center of the error function, and $\tau$ a measure for the width of the error function. In Fig. 3b, we show the parameter values of $t_0$ and $\tau$ from fits to the experimental data as a point with an error bar representing the uncertainty of the fit and as a colored bar, respectively. Additionally, we show a comparison of the fits to experimental data and simulations in **Fig. S17**. The fitted values for $t_0$ and $\tau$ are listed in **Table S3**.

**Note 3, Conrotatory Hydrogen Motion:** The evolution of the excited state wavepacket of αTP within the first 80 fs in conrotatory movement of the hydrogens around the ($C_3$) and ($C_4$) carbons is depicted in **Fig. S12** (coordinate defined in the caption and in Ref. 23). For comparison, also the projection onto the complementary disrotatory coordinate is shown. The wavepacket motion has significantly stronger conrotatory than disrotatory character. Additionally, the conrotatory motion is Franck-Condon active. However, its amplitude is only on the order of 1 degree, which points to the presence of the ($C_3$-$C_4$) σ bond constraining $CH_2$ rotation. Moreover, the timescale of the motion does not coincide with internal conversion or ($C_3$-$C_4$) bond dissociation.

**Note 4, Conrotatory Planarization and Deplanarization Motions:** The amplitudes of the individual planarization motions (green and purple curves in **Fig. 4b**) show close to identical amplitudes, whereas the corresponding deplanarization motions in **Fig. 4c** show a stronger amplitude for the deplanarization motion involving the methyl group. The behavior is connected to the much larger moment of inertia of the isopropyl substituent compared to the methyl substituent. Thus, the planarization motion around the ($C_3$=$C_1$) double bond is dominated by motion of the ($C_3$) hydrogens, whereas the methyl substituent is more strongly involved in the corresponding motion around the ($C_4$=$C_5$) double bond. This observation is in line with our interpretation of the early onset of the ɣ as being dominated by out-of-plane motion of the methyl group.

**Note 5, Structural Signatures of αTP Rotamers:** αTP exhibits a total of three ground state conformers, which differ by the rotation of the isopropyl group around the bond connecting it to the ring (M, P, and T in **Figs. S11** and **S18**). **Figure S18** shows the simulated PDFs as well as the carbon coordination sphere contributions for each of the rotamers compared to experiment. The most significant difference between the experimental and simulated PDFs in **Fig. S18a** is the magnitude of the minimum at 3.5 Å, which is connected to the bimodal distribution in the third carbon-carbon coordination sphere. The reasons for these differences are differences in orientation of the ($C_7$) and ($C_8$) carbon atoms with respect to the ring plane. In the M and P rotamers, one of the carbons is oriented in the ring plane and the other one in approx. 120° angle to the ring plane. In the T-rotamer, both carbons are oriented with an absolute angle of approx. 120° degree with respect to the ring plane. Thus, for the M and P rotamers, only third coordination sphere distances related to one of the two carbons fall in the center of the third coordination sphere distance range. For the T rotamer, distances related to both carbons fall into the center of the third coordination sphere range. This results in a deeper minimum for M and P rotamers than for the T rotamer. This feature can be reviewed from **Fig. S18b**, where two small peaks display in the center region of the 3rd coordination sphere in rotamer T. The fact that the minimum depth of the experimental PDF falls in between the m/p and t rotamer values suggests that our sample consisted of a thermal equilibrium of the three rotamers.

In addition to the static ground state PDFs, **Fig. S19** shows the computed time-dependent ΔPDF for each rotamer from the excited-state dynamics and compared directly to the measured signal. Panels a, b and c reflect the signals of M, P, and T, respectively. Panel d shows a lineout of the signals at a delay around 550 fs among the three rotamer and the measured signal. In general, the rotamers undergo similar dynamics and the amplitudes of the peaks and troughs match well. According to panel d, ΔPDFs of the rotamer m/p are almost identical and the main difference is from t at the pair distance between 3 and 5 Å. However, the difference from the different rotamers cannot be distinguished from the experimental measurement.

**FIGURES**

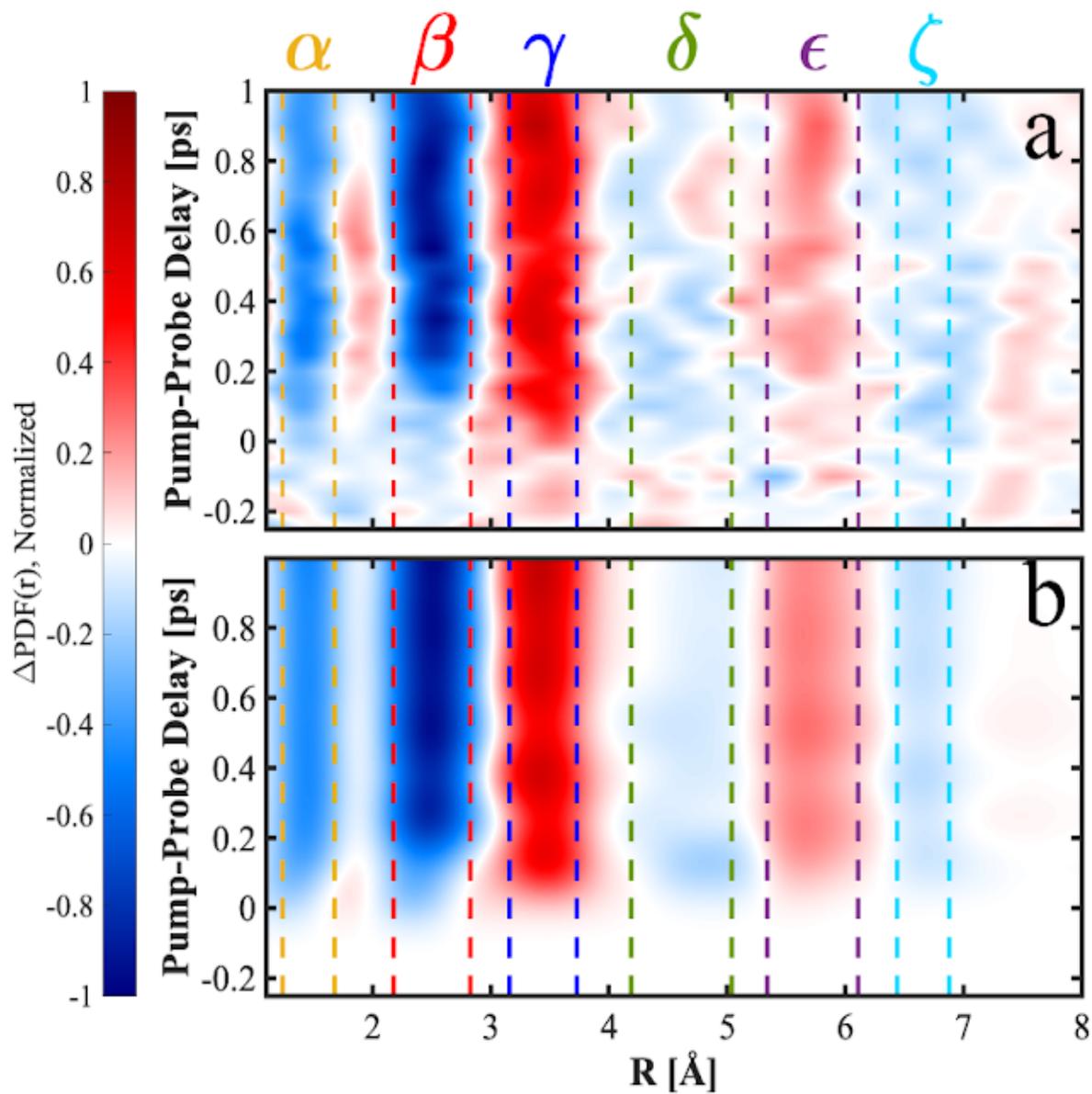

**Figure S1.** Experimental (a) and simulated (b) ΔPDF for the first 1 ps of αTP Photoinduced Ring-opening Dynamics.

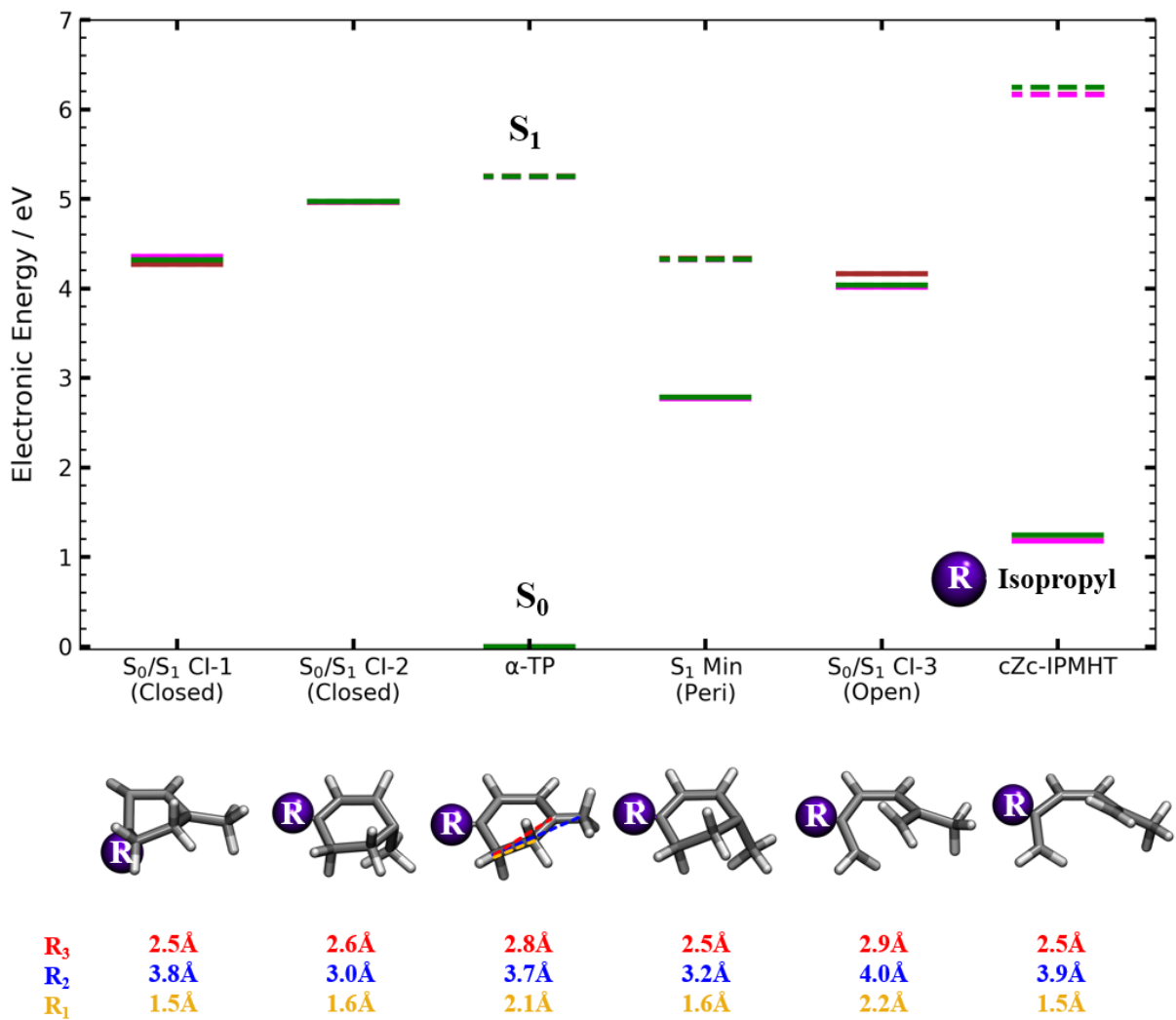

**Figure S2. Critical Points Along Nonradiative Relaxation Pathways in αTP.** The energies are relative to each rotamer's respective $S_0$ minimum ground-state energy (rotamer M (red), rotamer P (pink), and rotamer T (green)). Geometries shown in the bottom of the figure with the purple spheres representing the location of the isopropyl are computed at the α (0.82)-SA2-CAS(6,4)-SCF/6-31G* level of theory. See table S2 for Energies, Cartesian Coordinates, and CI Vectors.

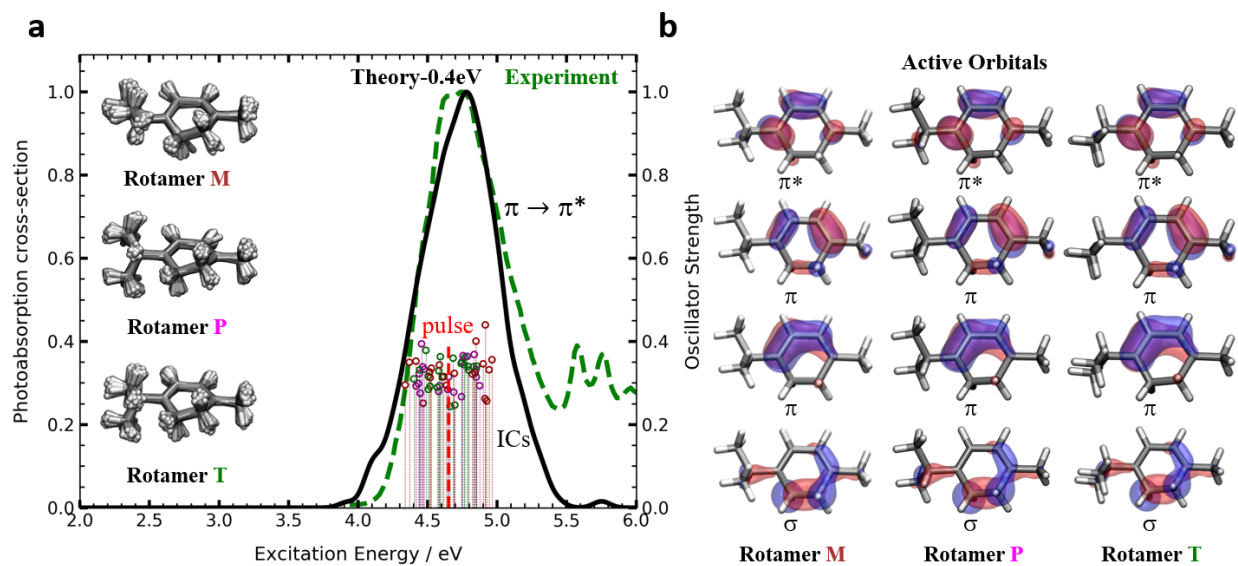

**Figure S3. UV Electronic Absorption Spectrum of αTP.** The computed UV electronic absorption spectrum (black) was generated from 600 initial conditions sampled from a ground-state harmonic Wigner distribution and compared against experiment (green). The AIMS dynamics simulations used 20 initial conditions for each rotamer (M (red), P (pink), T (green)). In panel a, the energy and oscillator strength for each of the initial conditions (randomly sampled with the restriction that they were within 0.3 eV of the pump pulse energy used in the UED experiment) are shown with red/pink/green vertical lines for the M/P/T initial conditions, respectively. The inset shows the starting geometries for each rotamer examined in this study. Panel b shows the SA-CASSCF natural molecular orbitals of the (6,4) active space at the $S_0$ minima of the three different rotamers. Blue and red correspond to 0.05 and -0.05 $e^-/Å^3$ isovalues, respectively.

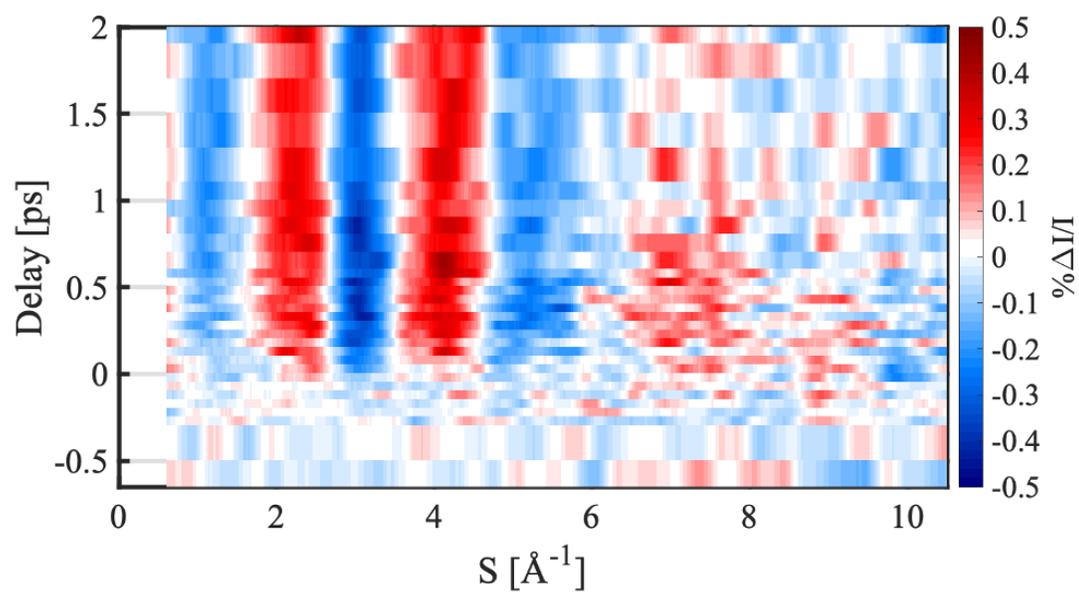

**Figure S4. Experimental time-dependent percentage difference diffraction signal.**

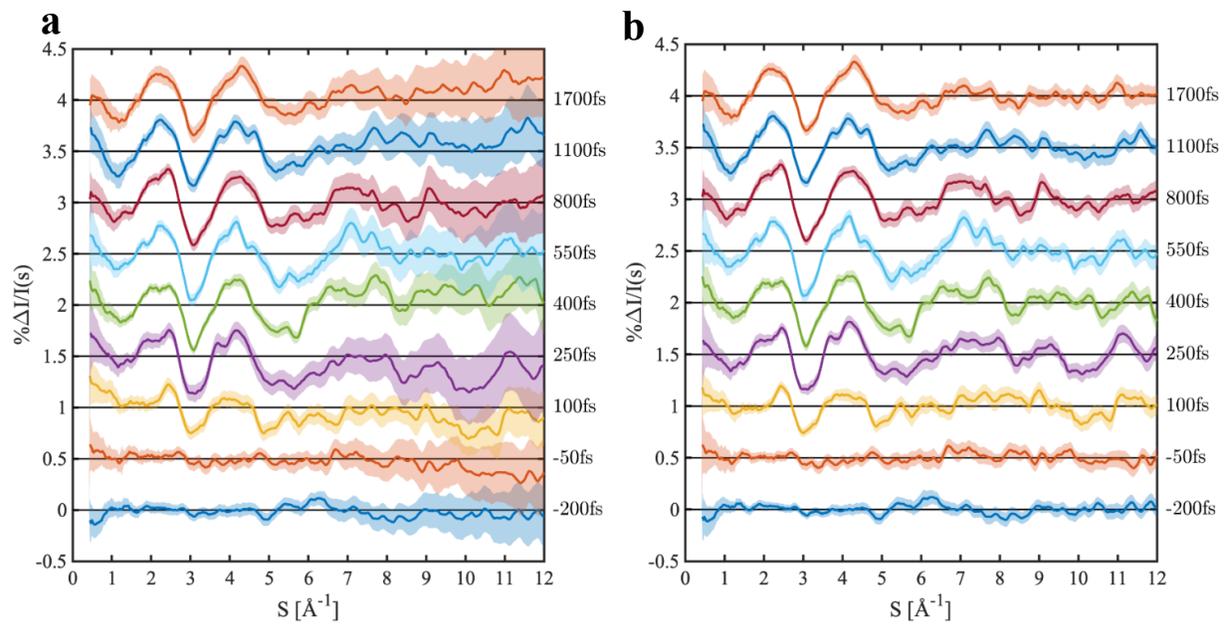

**Figure S5. Low-order polynomial background removal of the diffraction signal.** Panel **a:** Diffraction percentage difference signal (%ΔI/I) at different delays is plotted as colored lines. The corresponding shaded areas reflect the error bar (1 standard deviation). Panel **b:** Analogous plot to panel **a**, but after the low-order polynomial background removal process. The black lines indicate the baseline of each curve. The delay information is labeled on the right.

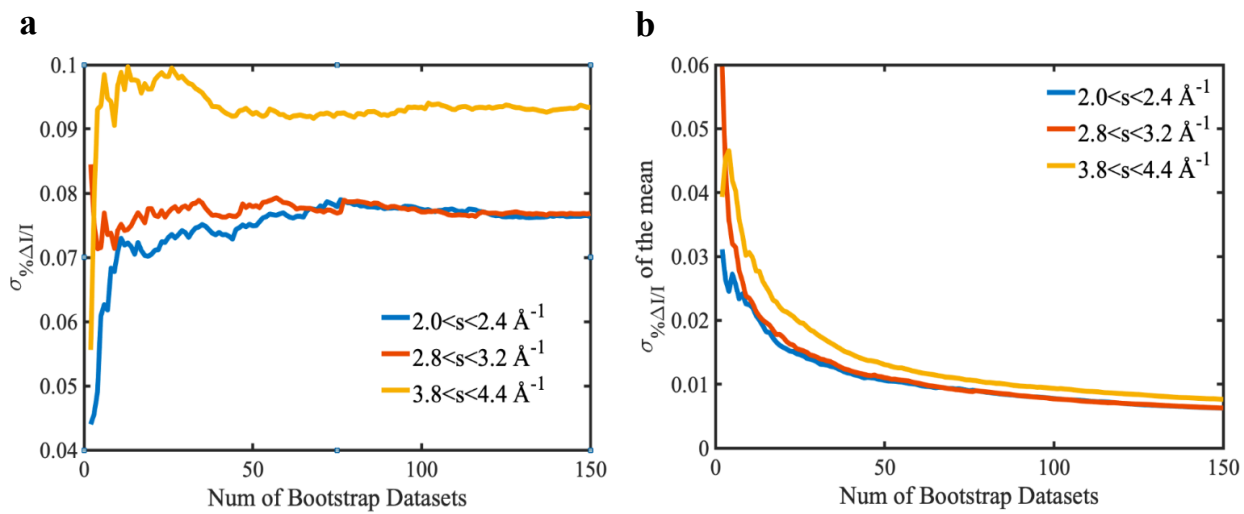

**Figure S6. Statistical uncertainty development of the bootstrapping analysis.** The uncertainty as a function of the number of bootstrapping dataset used in the analysis. Panel **a** reflects one standard deviation of the PDF at several different s regions, whereas panel **b** indicates the s standard deviation of the mean. The uncertainty undergoes a saturation after 150 bootstrapping datasets.

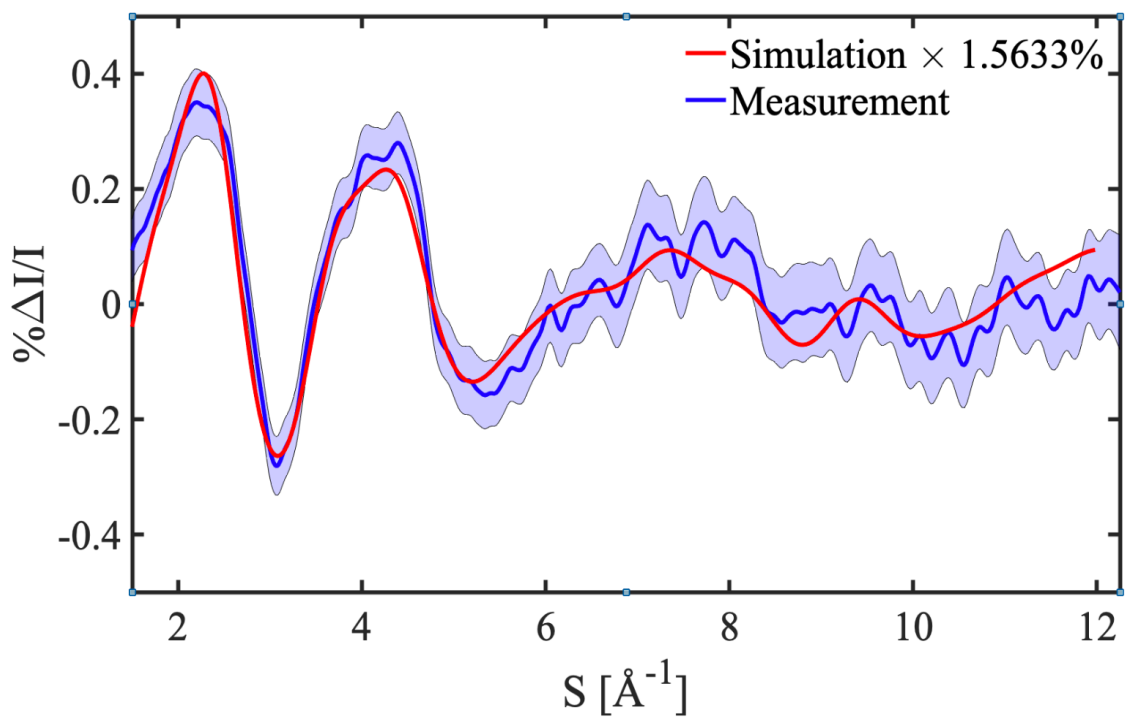

**Figure S7. Experimental and simulated percentage difference of the diffraction signal.** In this figure, two curve plots show the experimental signal and simulated percentage difference signals at 1 ps delay. The simulation is scaled by a factor of 0.0156 to match the amplitude of the experiment. This suggests an excitation ratio of $\gamma$ = 1.56%.

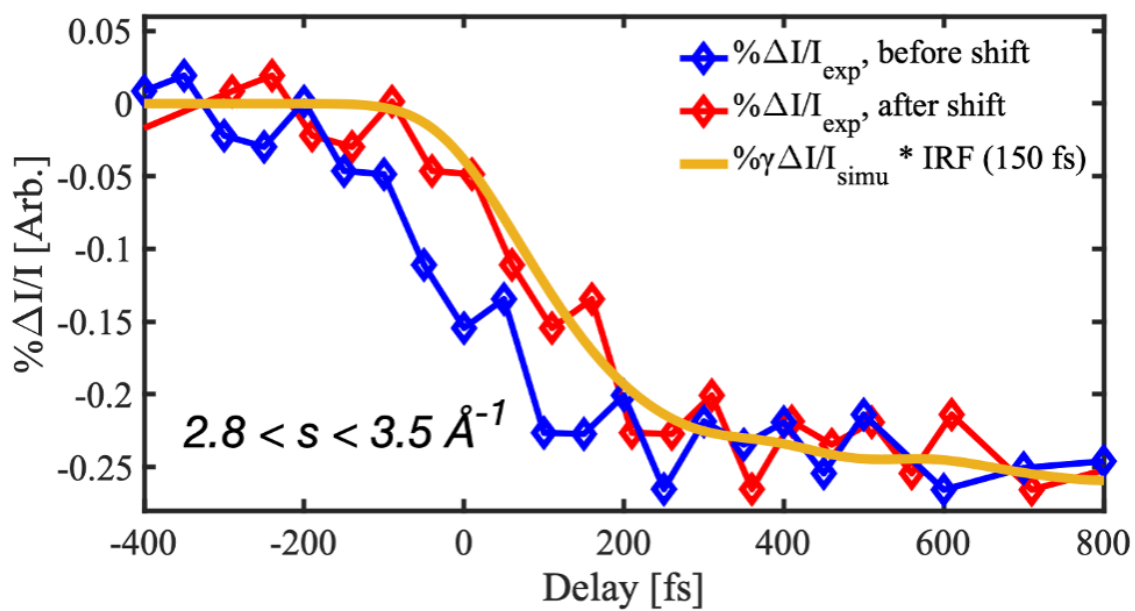

**Figure S8. Determination of the exact time-zero of the experiment.** Both the simulation and experimental data are integrated between 2.8 and 3.5 Å$^{-1}$. The simulation is convolved with a Gaussian function of FWHM of 150 fs.

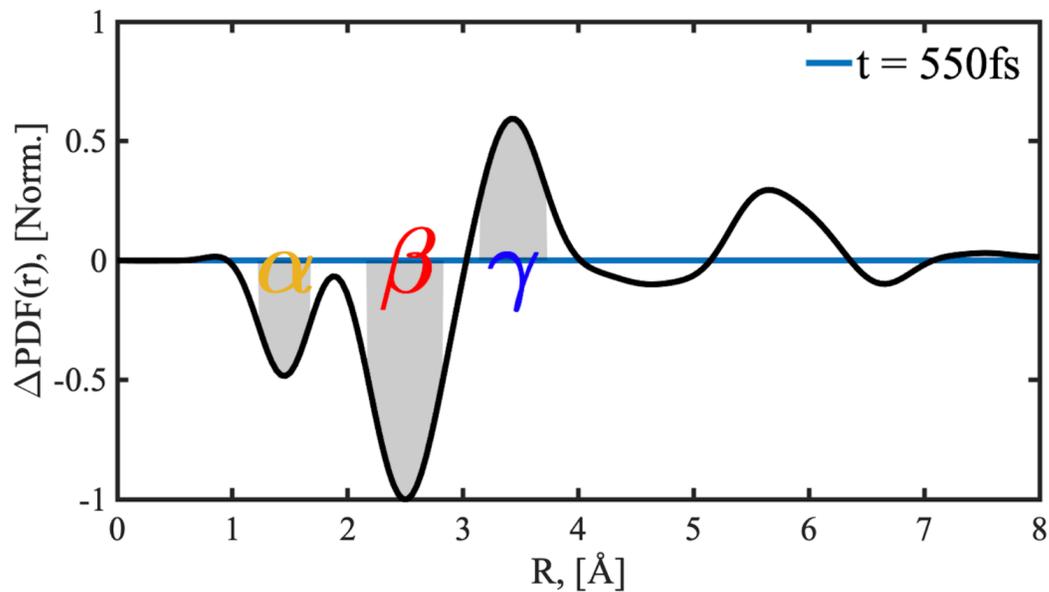

**Figure S9. Determination of the areas corresponding to different integration ranges of Fig. 3.** The line in the figure shows the averaged ΔPDF(r) at T = 550 fs of all the three rotamers, whereas the gray plots indicated the areas' limits of interest corresponding to the carbon coordination spheres. The areas are chosen according to the FWHM of each peak or trough of the simulated ΔPDF(r).

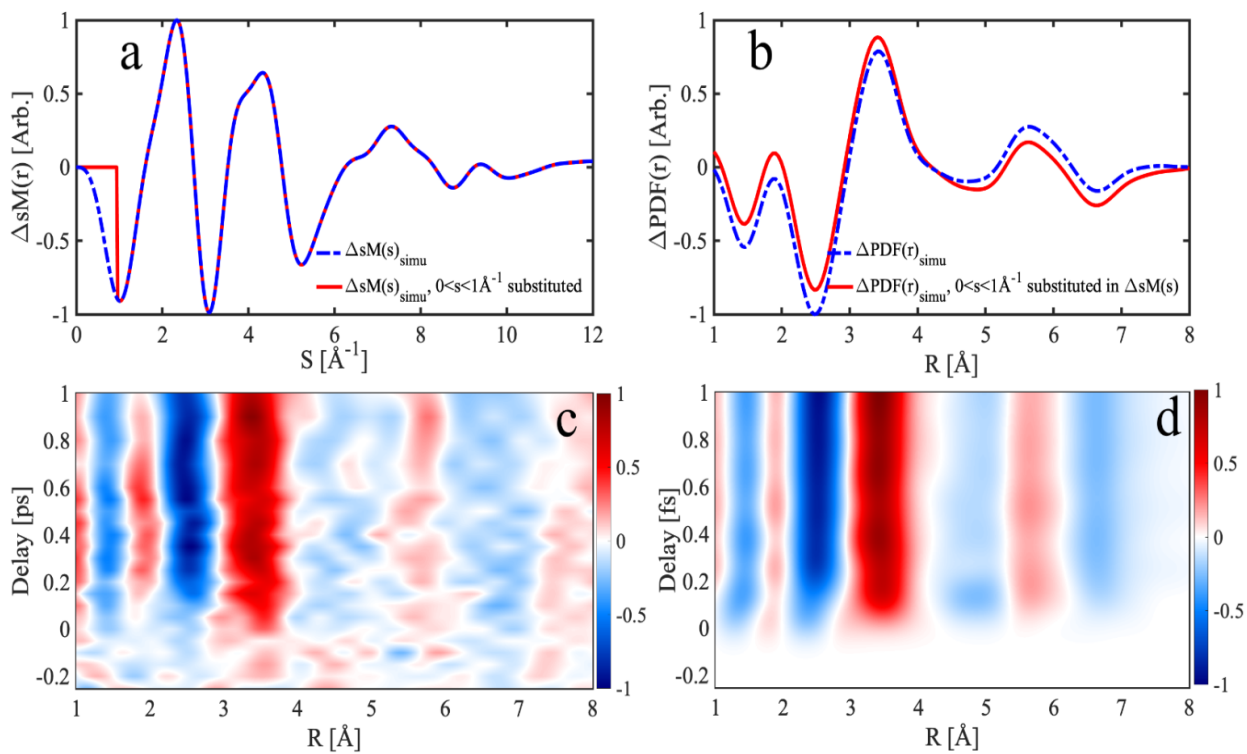

**Figure S10. Investigation of the low s signal in the calculation of the pair distribution function.** In this figure, we show the influence of the low s signal in calculation of the pair distribution function. Panel **a** shows the simulated ΔsM(s) signal around 500 fs. The blue dashed curve reflects the simulation, whereas the red curve reflects the low s signal substituted by zeros. Panel **b** shows the ΔPDF(r) based on the signals in panel **a**. Panels **c** and **d** plot the time-dependent ΔPDF(r) of experiment and simulation with the low s signal replaced by zeros.

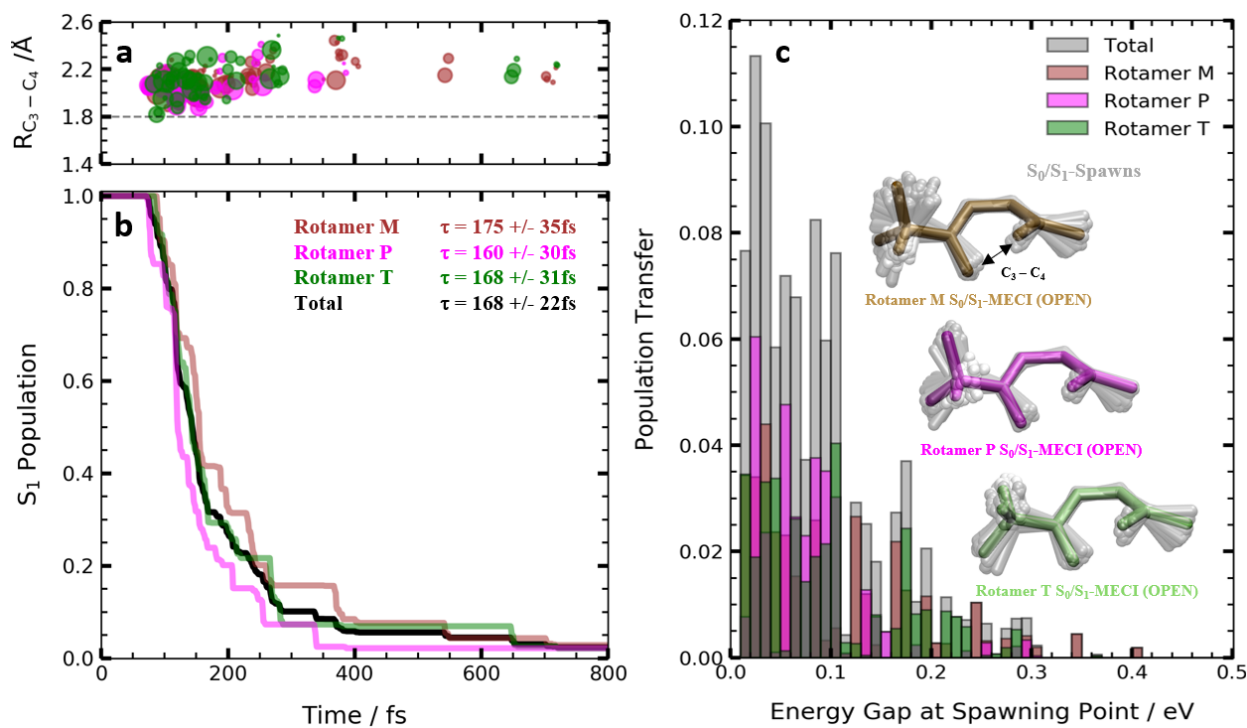

**Figure S11. a)** The $C_3$-$C_4$ bond distance vs spawning time at the spawning geometries from the AIMS trajectories for rotamers M (red), P (pink), and T (green). The circle radius is proportional to the population transferred during the spawning event. The population transfer is defined as the total population transferred to a child TBF from the beginning of coupled propagation until the child TBF becomes completely uncoupled (off-diagonal elements in the Hamiltonian become small). **b)** The population of the wavepacket on the $S_1$ adiabat for the first 800 fs of the photodynamics of αTP. Decay constants from single-exponential fits along with bootstrapped errors are shown in the inset. **c)** Histogram of the population transfer vs the energy gap of all $S_1/S_0$ spawning geometries from the AIMS simulation for all three rotamers. Spawning geometries superimposed on their respective ring-open MECI are shown in the inset with all hydrogens (except for the isopropyl) removed for clarity.

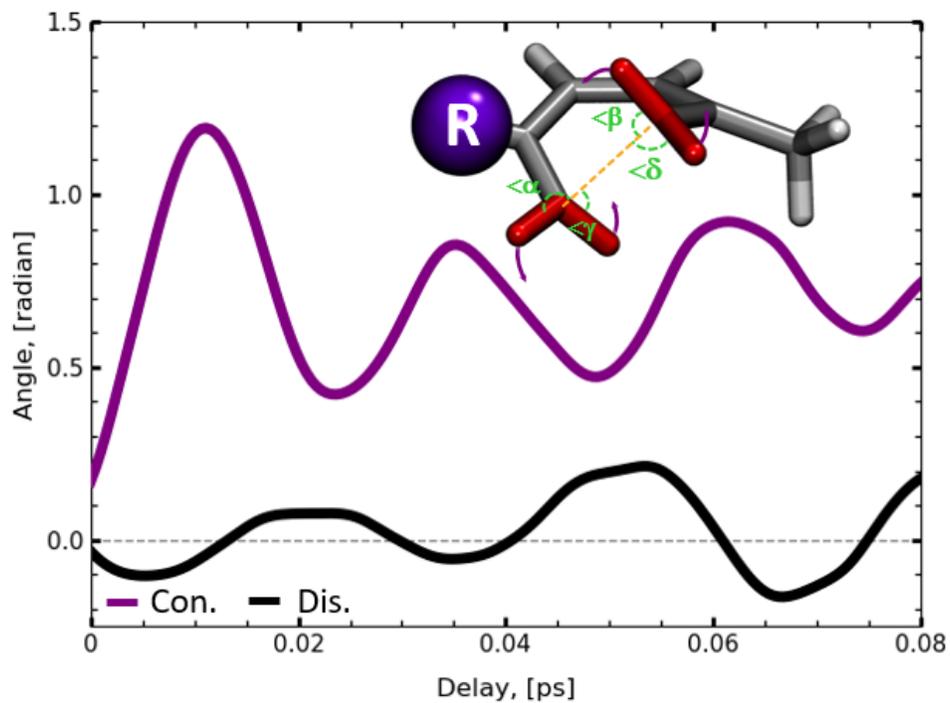

**Figure S12. The Averaged Conrotatory and Disrotatory Angles of αTP on S$_1$.** The direct average for the conrotatory and disrotatory angles from all 60 ICs in the first 80fs of propagation on S1. The conrotatory angle increased from around 0 to >1.0 radians while the disrotatory angle remained low. The conrotatory and disrotatory angles are defined as <α + <δ − <β − <γ and <α + <β − <δ − <γ, respectively, and are shown in the inset. The purple sphere represents the isopropyl group without any specific orientation.

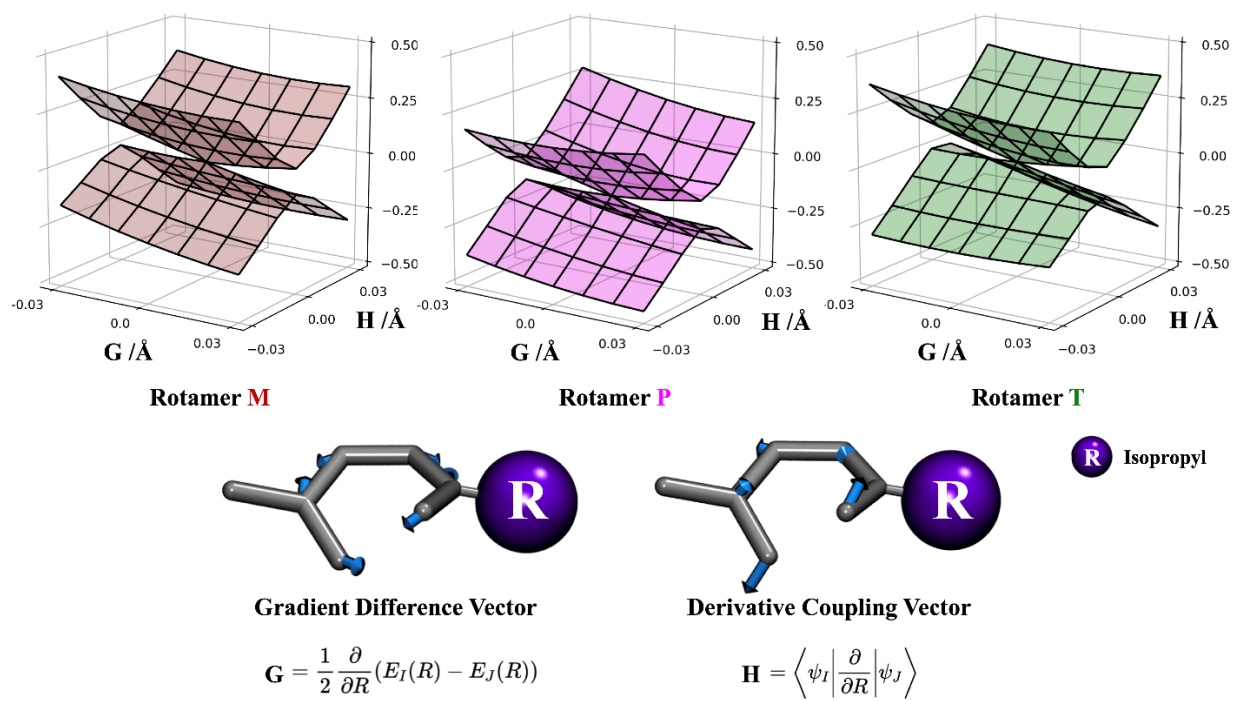

**Figure S13. The Branching Space at the Ring-Open $S_0/S_1$ MECI for the M, P, and T Rotamers.** The branching space at the ring-open $S_0/S_1$ MECI geometry for rotamers M (red), P (pink), T (green). The gradient-difference vector (left) and coupling vector (right) are shown on the ring-open MECI geometry with blue arrows. The purple sphere represents the isopropyl group without any specific orientation. All rotamers show a peak-like topography at the ring-open MECI point like CHD and αPH.

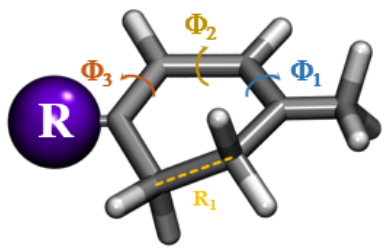

| Conformer | $R_1$ (Å) | $|\Phi_1|$ (°) | $|\Phi_2|$ (°) | $|\Phi_3|$ (°) |
|---|---|---|---|---|
| αTP | ≤ 1.8 | ≤ 80 | ≤ 80 | ≤ 80 |
| cZc-IPMHT | > 1.8 | ≤ 80 | ≤ 80 | ≤ 80 |
| cZt/tZc-IPMHT | > 1.8 | ≤ 80 or ≥ 100 | ≤ 80 | ≥ 100 or ≤ 80 |
| tZt-IPMHT | > 1.8 | ≥ 100 | ≤ 80 | ≥ 100 |

**Figure S14. Classifying αTP and it IPMHT Isomers on $S_0$.** Three angles and one distance are utilized to define the geometries of the isomers. In addition to αTP, three different ring-opening isomers are identified, cZc, cZt/tZc, and tZt-IPMHTs. As shown by the molecular cartoon in the top, three dihedral angles, $\Phi_1$, $\Phi_2$, $\Phi_3$ across the CHD ring structure and distance $R_1$ between $C_3$-$C_4$. In the table, each row corresponds to the four classification criteria used to bin geometries from the ground state TBF trajectories..

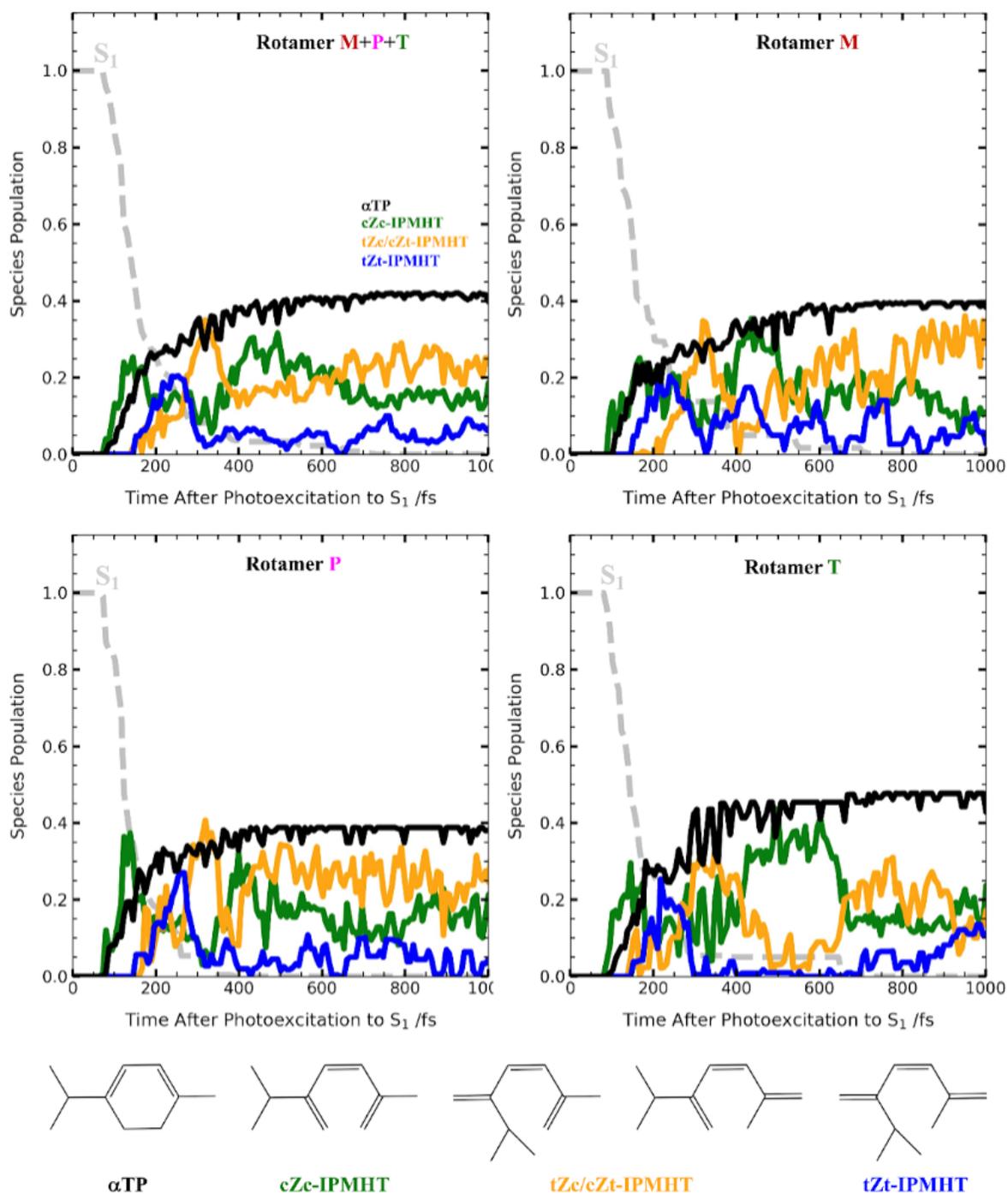

**Figure S15. Population of αTP and ring-opening product isomers on the S$_0$ after excitation.** The percentage of the ground-state population binned into IPMHT isomers from optimized ground-state geometries via torsional angles $\Phi_1$, $\Phi_2$, $\Phi_3$, and $R_1$ (see supplementary Fig. S15). Time zero corresponds to the spawn time of each TBF to S$_0$. Panels **a** to **d** show the relative population of the different product isomers as a function of time as well as the excited state population. Panel **a** reflect the population of all the three-ground state rotamers, whereas panels **b** to **d** showcase the populations from the individual ground state rotamers. The cartoon diagrams of the geometries are depicted in the bottom of the figure.

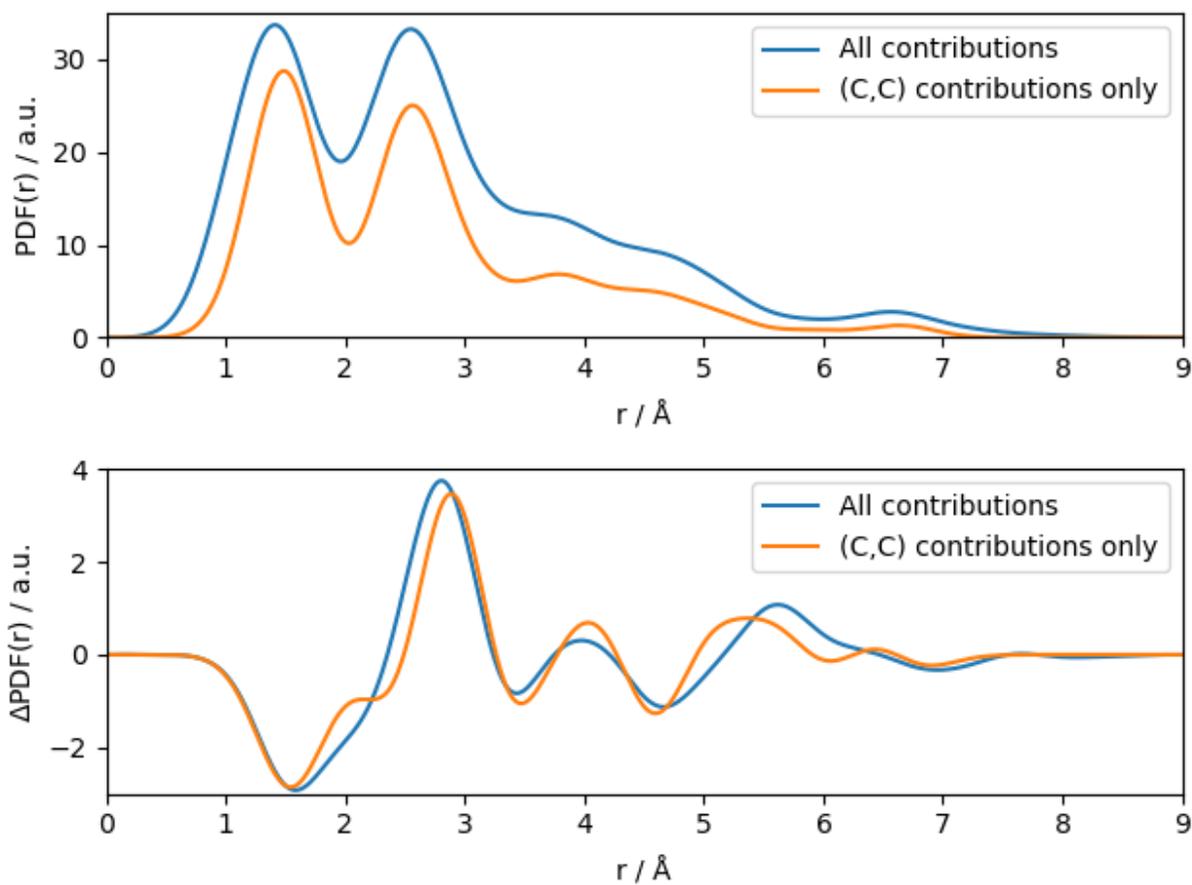

**Fig. S16. Contributions from C-H and H-H distances to pair distribution functions (PDF) and difference pair distribution functions (ΔPDF).** Top: Comparison of simulated PDFs of αTP including all contributions and (C,C) contributions only. Bottom: Analogous simulations of ΔPDF of the cZc photoproduct isomer of αTP.

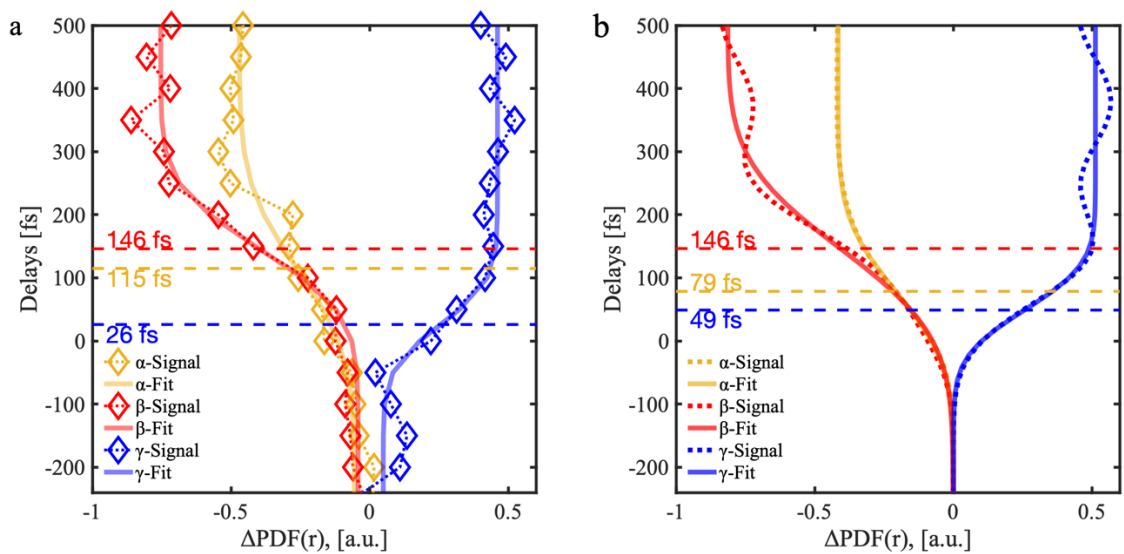

**Figure S17. Investigation of signal onset time at different pair distances.** In this figure, we show the extracted signal onset time of α, β, and γ region for both the experiment (panel a) and simulation (panel b). We fit the signal with an error function at each region, respectively. The horizontal dashed lines in both panels indicate the onset times respect to T0 and relevant values are labeled.

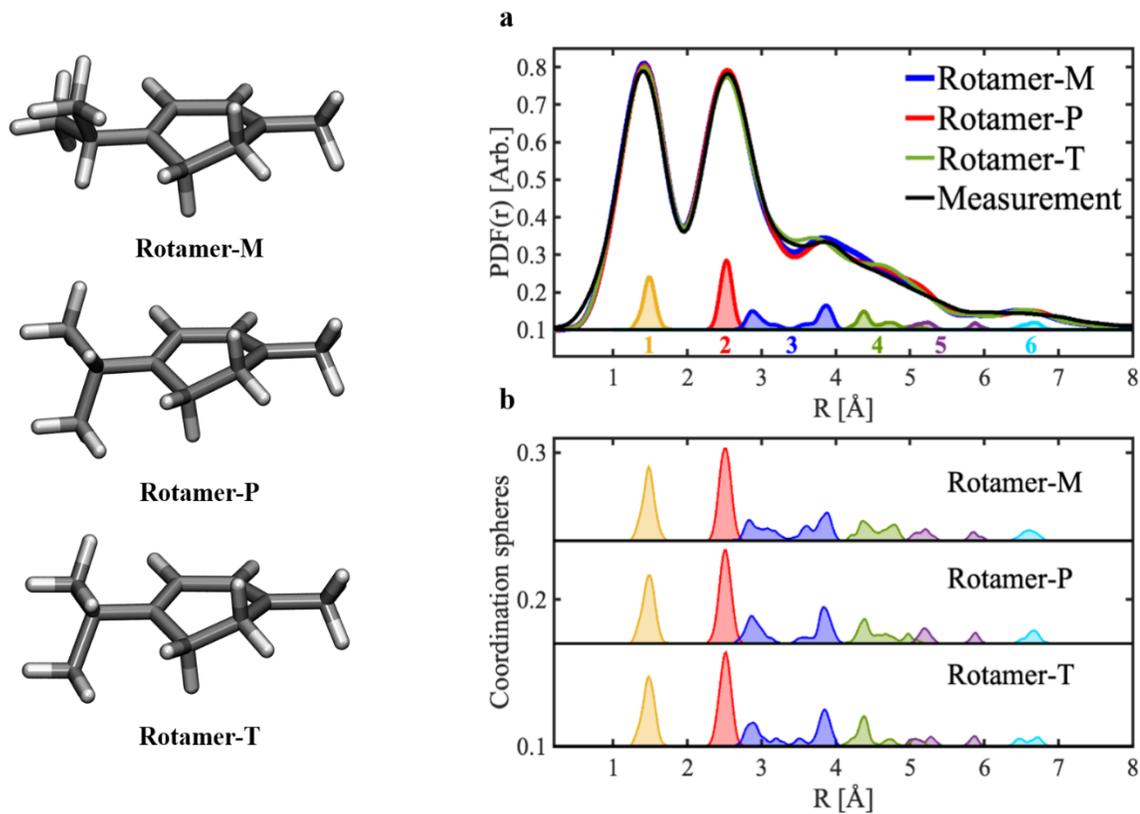

**Figure S18. Pair distribution functions of αTP ground state rotamers.** Panel **a** plots both the experimental and simulated ground state pair distribution functions. The simulated PDF(r) include all the three rotamers. The difference of PDF(r) between rotamer M and P is rather small and only rotamer T displaces certain amount of difference at the 3rd and 4th coordination shells. The colored shaded areas underneath the PDF curves reflect the relevant contribution from carbon-carbon atomic pairs in each different carbon coordination spheres from all the three rotamers. The shading plots in panel **b** reflect the contribution of carbon coordination spheres from each individual rotamers.

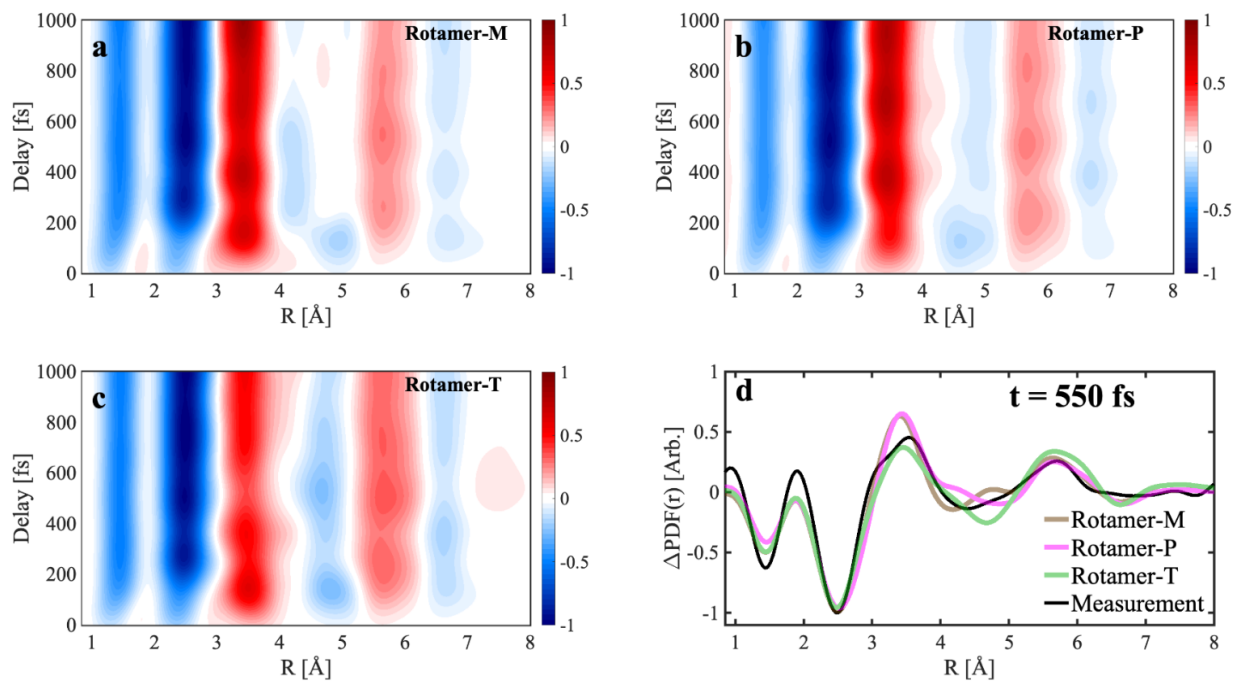

**Figure S19. ΔPDF(r, t) from three rotamers of α-terpinene.** Panels **a**, **b**, and **c** reflect the time-dependent ΔPDFs from the AIMS trajectories of rotamers M, P and T, respectively. Signal in each panel is averaged from 20 initial geometries. Panel **d** shows the ΔPDFs at t = 550 fs of the three rotamers.

# TABLES

**Table S1. Computed Critical Points Raw Data.** Cartesian coordinates, CI vectors, and Energies for All Critical Points in the Three Rotamers of Figure S2.

## ROTAMER M

### S$_0$ Minimum

```
26
C    0.7041856002  -0.2440646117   0.0335001222
C    2.2163531025  -0.3113246546   0.1226795910
C    0.1010709204   1.0165034948  -0.5509225670
C   -1.3016751784   1.3077789056  -0.0155723140
C   -2.1677780262   0.0691348265   0.0309644201
C   -0.1386342130  -1.2534965556   0.3613420539
C    2.7301827610   0.4775620320   1.3399906120
C    2.7866108940  -1.7322905187   0.1365025521
C   -1.5767606644  -1.1124619112   0.2336379563
C   -3.6562565878   0.2464252513  -0.0686530446
H    2.6042535290   0.1831648254  -0.7682751805
H    0.7446079539   1.8721258102  -0.3722978827
H    0.0533671775   0.8946852447  -1.6345960159
H   -1.7724106198   2.0800474375  -0.6170890433
H   -1.2301548867   1.7097788314   0.9967438419
H    0.2379809159  -2.1993264234   0.7044727909
H    2.3970227405   0.0116913846   2.2624466730
H    2.3720982379   1.5017728678   1.3379455454
H    3.8160206922   0.5074071909   1.3493874776
H    2.4394682168  -2.3074987647  -0.7151582953
H    2.5087329166  -2.2662125309   1.0398876437
H    3.8713274598  -1.6995786564   0.1004538756
H   -2.1784874898  -2.0014976711   0.3273695690
H   -4.0152779407   0.9359711165   0.6933157862
H   -4.1810077554  -0.6942253965   0.0540083798
H   -3.9378397558   0.6679284756  -1.0310845477
```

State Energies and CI Vectors with determinants (X-closed, A-alpha open, B-beta open)

State 1 energy: -387.96735370474971 [Ha]
        0.97974343242992  X36 X37 X38
    - 0.12034485841768  X36 X37 X39
        0.08065554023009  X36 X37 A38 B39
        0.08065554023009  X36 X37 B38 A39
    - 0.07418353896875  X36 X38 X39
    - 0.04594692785808  X36 A37 B38 X39
    - 0.04594692785808  X36 B37 A38 X39
        0.03318103195400  A36 X37 B38 X39
        0.03318103195400  B36 X37 A38 X39

State 2 energy: -387.77423395415337 [Ha]
    - 0.67887732464152  X36 X37 A38 B39
    - 0.67887732464152  X36 X37 B38 A39
        0.13110784565862  X36 A37 X38 B39
        0.13110784565862  X36 B37 X38 A39
        0.12114413996907  X36 X37 X39
        0.11430412820650  X36 X37 X38
    - 0.07330782429894  X36 X38 X39
    - 0.05642292767479  X36 A37 B38 X39
    - 0.05642292767479  X36 B37 A38 X39
    - 0.03090676560869  A36 X37 X38 B39
    - 0.03090676560869  B36 X37 X38 A39
    - 0.02542297688135  A36 X37 B38 X39
    - 0.02542297688135  B36 X37 A38 X39
        0.02412600685164  A36 B37 X38 X39
        0.02412600685164  B36 A37 X38 X39

## MECI-1 (Closed)

```
26
C    0.5667045484  -0.2243628089  -0.3643548941
C    1.2382695187  -0.4664227637  -1.6972233264
C    0.3810431305   1.1926408909   0.0957515958
C   -1.1420324862   1.2441006739   0.3126278631
C   -1.7275676025  -0.0403069361  -0.2546976235
C    0.1636534254  -1.2728103365   0.5735374773
C    2.7606387212  -0.3574101766  -1.4561616866
C    0.8853522451  -1.7923067391  -2.3662032653
C   -1.1563134984  -1.2604064320   0.1407948312
C   -3.0679878481   0.0281188268  -0.9305240991
H    0.9622677955   0.3412763198  -2.3689749892
H    0.9373088374   1.3460620203   1.0167781582
H    0.7070821357   1.9152118813  -0.6463173407
H   -1.5575957820   2.1052422638  -0.1979521595
H   -1.3959866501   1.3433774923   1.3679182506
H    0.7018229590  -2.1965124870   0.6644562124
H    3.1067981741  -1.1424252577  -0.7920752809
H    3.0353010284   0.5987485790  -1.0228625297
H    3.2869460209  -0.4557705371  -2.3999827921
H   -0.1842708042  -1.8750416953  -2.5204296118
H    1.2098096711  -2.6388540293  -1.7734325695
H    1.3709836620  -1.8522837609  -3.3353669514
H   -1.7252804271  -2.1705509398  -0.0235396645
H   -3.7969667291   0.5515158301  -0.3095360356
H   -3.4613833244  -0.9609895609  -1.1429029052
H   -3.0244263068   0.5689262613  -1.8749373727
```

State Energies and CI Vectors with determinants (X-closed, A-alpha open, B-beta open)

State 1 energy: -387.81055400366597 [Ha]
          0.91043962236385  X36 X37 X38
          0.26400508628763  X36 X37 A38 B39
          0.26400508628763  X36 X37 B38 A39
        - 0.09834423722041  X36 A37 X38 B39
        - 0.09834423722041  X36 B37 X38 A39
        - 0.08551352887311  X36 X38 X39
        - 0.04085332005156  X37 X38 X39
          0.03888616986593  X36 A37 B38 X39
          0.03888616986593  X36 B37 A38 X39

State 2 energy: -387.81055388607899 [Ha]
        - 0.64395041906167  X36 X37 A38 B39
        - 0.64395041906167  X36 X37 B38 A39
          0.37779296099478  X36 X37 X38
        - 0.11346410854109  X36 A37 B38 X39
        - 0.11346410854109  X36 B37 A38 X39
        - 0.03573239697975  X36 X38 X39

## MECI-2 (Closed)

```
26
C    -0.1615322911    0.5145425888   -0.6118593811
H     0.0418735091    0.5233931219   -1.6816495207
C    -1.2795121996    1.5660729028   -0.3105364089
C    -0.5060997782   -0.9209572407   -0.2954414116
H     1.4598795398   -1.5259164395   -0.7290001955
C     0.3939793125   -3.3278148982   -0.3443845506
C    -2.6995576601    1.0300202544   -0.4673864086
C    -2.7848979236   -0.3053722000   -0.0517957876
C    -1.7881927208   -1.3399214145   -0.2537292648
H    -3.6327971208   -0.6584589573    0.5353119532
H     0.7805771544    0.7789404292   -0.1381404601
C     1.1861654706   -1.6652283787    1.3842476015
H    -1.1143462543    2.4313877237   -0.9429290922
H    -1.1582836608    1.9149012258    0.7120762683
C     0.6707981023   -1.8501903682   -0.0545880209
H    -0.3086965048   -3.7559578465    0.3625347906
H     1.3165011139   -3.8936570455   -0.2662375603
H    -0.0024322589   -3.4709824217   -1.3437855600
H     2.0889895187   -2.2475952877    1.5353379646
H     0.4439491027   -1.9991309190    2.1020455289
H     1.4204321311   -0.6296809130    1.6022456505
C    -2.7896664837    0.4923477011   -2.0447022047
H    -2.0362370550   -2.3675346688   -0.0591710873
H    -3.0930869804    1.4480993791   -2.4596783934
H    -3.6065616522   -0.1902274430   -2.2539401574
H    -1.8973466250    0.1125569520   -2.5308444390
```

State Energies and CI Vectors with determinants (X-closed, A-alpha open, B-beta open)

State 1 energy: -387.78497668589318 [Ha]
```
        0.60216159069175   X36 X37 A38 B39
        0.60216159069175   X36 X37 B38 A39
       -0.49648778130800   X36 X37 X38
        0.09109982784433   X36 A37 B38 X39
        0.09109982784433   X36 B37 A38 X39
        0.05696871502601   X36 X38 X39
        0.05114125275460   X36 A37 X38 B39
        0.05114125275460   X36 B37 X38 A39
        0.03518385543177   A36 X37 B38 X39
        0.03518385543177   B36 X37 A38 X39
        0.02618392304446   X37 X38 X39
```

State 2 energy: -387.78497664401840 [Ha]
```
        0.84539587579105   X36 X37 X38
        0.35344617447313   X36 X37 A38 B39
        0.35344617447313   X36 X37 B38 A39
       -0.09428724612450   X36 A37 X38 B39
       -0.09428724612450   X36 B37 X38 A39
       -0.09241436097864   X36 X38 X39
       -0.05498803392665   X37 X38 X39
        0.05249224553938   X36 A37 B38 X39
        0.05249224553938   X36 B37 A38 X39
```

## S₁ Minimum (Pericyclic)

```
26
C     0.5517386695   -0.0659086046    0.2128944271
C     2.0558143529   -0.0881504139    0.3919634194
C    -0.0076373576    0.5153876263   -1.0246366000
C    -1.2447241457    1.4635634582   -0.6641456700
C    -2.2629164830    0.7031461799    0.1196548367
C    -0.3248084163   -0.4466734195    1.2552154781
C     2.5237348283    1.2777689000    0.9309562185
C     2.5781397999   -1.2317263280    1.2685888753
C    -1.6276435783    0.0191142568    1.2439640324
C    -3.0193258682   -0.3292017463   -0.7059854339
H     2.4854755613   -0.2128782010   -0.5975138181
H     0.7653749113    1.0207865673   -1.5983248582
H    -0.4187687344   -0.2711496576   -1.6534367664
H    -1.6499007375    1.8452199045   -1.5954602666
H    -0.8436038186    2.3141745790   -0.1169104850
H     0.0748569828   -0.8636551090    2.1615662239
H     2.1235375821    1.4579974295    1.9233240767
H     2.2075279552    2.0916729098    0.2876085623
H     3.6069098493    1.3028672061    0.9933608059
H     2.2026963415   -2.1935836748    0.9351048359
H     2.3054745840   -1.1048498483    2.3104557447
H     3.6616291195   -1.2618389533    1.2189349220
H    -2.1715372258   -0.1248908312    2.1698215527
H    -3.4988125455    0.1573198604   -1.5511033180
H    -3.8087896840   -0.8059599521   -0.1288991028
H    -2.4065961126   -1.1509068843   -1.1173750861
```

State Energies and CI Vectors with determinants (X-closed, A-alpha open, B-beta open)

State 1 energy: -387.86505913068225 [Ha]
```
          0.59768735217718  X36 X37 X38
        - 0.54433877369313  X36 X37 A38 B39
        - 0.54433877369313  X36 X37 B38 A39
        - 0.15313149374745  X36 X37 X39
        - 0.09384372573376  X36 A37 B38 X39
        - 0.09384372573376  X36 B37 A38 X39
        - 0.05355383035958  A36 X37 B38 X39
        - 0.05355383035958  B36 X37 A38 X39
        - 0.04099815836357  X36 X38 X39
        - 0.02195515537062  A36 X37 X38 B39
        - 0.02195515537062  B36 X37 X38 A39
```

State 2 energy: -387.80806814696808 [Ha]
```
        - 0.77045560501774  X36 X37 X38
        - 0.40362565538979  X36 X37 A38 B39
        - 0.40362565538979  X36 X37 B38 A39
          0.16542893925215  X36 X38 X39
          0.11650079551853  X36 A37 X38 B39
          0.11650079551853  X36 B37 X38 A39
        - 0.10706017227454  X36 X37 X39
        - 0.07317230853129  X36 A37 B38 X39
        - 0.07317230853129  X36 B37 A38 X39
          0.03577173788894  A36 X37 X38 B39
          0.03577173788894  B36 X37 X38 A39
          0.02337687662337  X37 X38 X39
```

## MECI-3 (Open)

```
26
C    0.7122216454   -0.2134316839   -0.0622418779
C    2.1763918521   -0.1385094836    0.3873839768
C    0.3247407478    0.4513869563   -1.2343782854
C   -1.3865081609    1.3865378071   -0.1343077348
C   -2.2222734371    0.2686040648    0.0604519090
C   -0.2074173699   -0.9691929140    0.7289323028
C    2.8031065610    1.2475783498    0.1579358126
C    2.4287728158   -0.5712462402    1.8387193693
C   -1.5912135651   -0.8451669186    0.6119608595
C   -3.6057483602    0.2000695226   -0.5437589775
H    2.7023433484   -0.8393486561   -0.2606107331
H    1.0365446570    1.0484701897   -1.7702962388
H   -0.4601266691    0.0370101712   -1.8367310714
H   -1.6759810298    2.1734590684   -0.8097356928
H   -0.7280070196    1.6905165068    0.6563118590
H    0.1759192758   -1.7232856003    1.3868235199
H    2.2693673863    2.0158230250    0.7096951244
H    2.8187703368    1.5299154956   -0.8854837308
H    3.8305559383    1.2435701808    0.5056567122
H    2.1899606268   -1.6136752850    2.0076996532
H    1.8495095539    0.0224921887    2.5381608659
H    3.4791397591   -0.4411728711    2.0777486487
H   -2.2026517582   -1.6638966423    0.9631210669
H   -4.1533542300    1.1290901696   -0.4153080769
H   -4.1863121659   -0.5831141331   -0.0649733702
H   -3.5939154175   -0.0206348639   -1.6117892535
```

State Energies and CI Vectors with determinants (X-closed, A-alpha open, B-beta open)

State 1 energy: -387.81427596550236 [Ha]

```
       - 0.97873485069403  X36 X37 X38
         0.14947748262954  X36 X38 X39
       - 0.08612535187171  X36 X37 X39
         0.08343484091266  X37 X38 X39
       - 0.04164674640115  X36 A37 X38 B39
       - 0.04164674640115  X36 B37 X38 A39
       - 0.02671692874132  X36 X37 A38 B39
       - 0.02671692874132  X36 X37 B38 A39
```

State 2 energy: -387.81427590997237 [Ha]

```
         0.70200190232702  X36 X37 A38 B39
         0.70200190232702  X36 X37 B38 A39
       - 0.07515855364472  X36 A37 B38 X39
       - 0.07515855364472  X36 B37 A38 X39
       - 0.03360985519600  X36 X37 X38
         0.02530995398208  X36 X38 X39
```

# Rotamer P

## S₀ Minimum

```
26
C     0.7091707280   -0.2555438461   -0.0427616682
C     2.2207470826   -0.3507441976    0.0333884262
C     0.1071549407    1.1124801343   -0.2934210912
C    -1.3012419141    1.2487562653    0.2876871841
C    -2.1647069659    0.0450542516   -0.0153461824
C    -0.1352618553   -1.3156882392   -0.0097687128
C     2.7433681538   -1.7085537847    0.5080168371
C     2.8678818256    0.0258812833   -1.3112275185
C    -1.5713639296   -1.1469609602   -0.1296501905
C    -3.6521093812    0.2432820572   -0.0863145691
H     2.5459701372    0.3912743506    0.7627824856
H     0.7462732633    1.8908362480    0.1133837084
H     0.0618061348    1.2840168196   -1.3695295771
H    -1.7662714341    2.1559395553   -0.0871927412
H    -1.2420035481    1.3602584052    1.3718046760
H     0.2375514927   -2.3174638530    0.0956257885
H     2.5303268720   -2.4926223142   -0.2121316982
H     2.3043245729   -1.9943738264    1.4578838163
H     3.8207129400   -1.6683031876    0.6361250809
H     2.5726230823    1.0168499492   -1.6390463315
H     2.5803022252   -0.6811291794   -2.0839230176
H     3.9508985633    0.0157648657   -1.2309959962
H    -2.1692866107   -2.0291485452   -0.2890972672
H    -4.0246449958    0.7021097024    0.8278295462
H    -4.1752696743   -0.6952662671   -0.2298518512
H    -3.9209517054    0.9092943125   -0.9032691367
```

State Energies and CI Vectors with determinants (X-closed, A-alpha open, B-beta open)

State 1 energy: -387.96735788549069 [Ha]
      0.98002633585842  X36 X37 X38
   - 0.11937427677254  X36 X37 X39
   - 0.07934868356525  X36 X37 A38 B39
   - 0.07934868356525  X36 X37 B38 A39
   - 0.07463464800500  X36 X38 X39
      0.04620587549916  X36 A37 B38 X39
      0.04620587549916  X36 B37 A38 X39
      0.03304566929251  A36 X37 B38 X39
      0.03304566929251  B36 X37 A38 X39

State 2 energy: -387.77461543619410 [Ha]
      0.67928231931451  X36 X37 A38 B39
      0.67928231931451  X36 X37 B38 A39
      0.13109102277168  X36 A37 X38 B39
      0.13109102277168  X36 B37 X38 A39
      0.11873156749063  X36 X37 X39
      0.11213666207736  X36 X37 X38
   - 0.07267368849282  X36 X38 X39
      0.05644755948318  X36 A37 B38 X39
      0.05644755948318  X36 B37 A38 X39
      0.03109732769080  A36 X37 X38 B39
      0.03109732769080  B36 X37 X38 A39
   - 0.02571761203968  A36 X37 B38 X39
   - 0.02571761203968  B36 X37 A38 X39
   - 0.02423023839256  A36 B37 X38 X39
   - 0.02423023839256  B36 A37 X38 X39

## MECI-1 (Closed)

```
26
C    0.5668895207  -0.2045668657  -0.3355255564
C    1.2878506029  -0.4875864723  -1.6387173276
C    0.3624113347   1.2013966511   0.1574872393
C   -1.1699391522   1.2529883884   0.3184548574
C   -1.7403349514  -0.0257826466  -0.2738069615
C    0.1368587458  -1.2491644722   0.5980990679
C    1.1495108847  -1.9314081423  -2.1287283961
C    0.9728267016   0.4829327693  -2.7873100831
C   -1.1698171735  -1.2437618003   0.1249139884
C   -3.0791091949   0.0387355508  -0.9528214494
H    2.3366016327  -0.3319765241  -1.3597278161
H    0.8770878338   1.3007257842   1.1091918735
H    0.7218902051   1.9660182977  -0.5193441947
H   -1.5658305518   2.1171368199  -0.2025396186
H   -1.4630848425   1.3478959859   1.3645219606
H    0.6713795688  -2.1730419972   0.6958236885
H    0.1180916437  -2.1479194413  -2.3885047288
H    1.4667981463  -2.6503592455  -1.3866806415
H    1.7551892722  -2.0753865242  -3.0173707796
H    1.1444017609   1.5185936971  -2.5249869101
H   -0.0613207026   0.3787505043  -3.0982324308
H    1.6063810849   0.2567645608  -3.6388940876
H   -1.7273138318  -2.1567907248  -0.0601753684
H   -3.8194893790   0.5219142279  -0.3130936907
H   -3.4491491363  -0.9509172703  -1.2002433852
H   -3.0479604644   0.6139625273  -1.8766509218
```

State Energies and CI Vectors with determinants (X-closed, A-alpha open, B-beta open)

State 1 energy: -387.80730212878512 [Ha]
```
      - 0.64877067061222  X36 X37 X38
      - 0.52155588841294  X36 X37 A38 B39
      - 0.52155588841294  X36 X37 B38 A39
        0.08757432986921  X36 A37 X38 B39
        0.08757432986921  X36 B37 X38 A39
      - 0.08518035901078  X36 A37 B38 X39
      - 0.08518035901078  X36 B37 A38 X39
        0.06167751187308  X36 X38 X39
        0.02894028764212  X37 X38 X39
```

State 2 energy: -387.80730203270952 [Ha]
```
      - 0.74174844405381  X36 X37 X38
        0.46037119331303  X36 X37 A38 B39
        0.46037119331303  X36 X37 B38 A39
        0.08586049032704  X36 A37 B38 X39
        0.08586049032704  X36 B37 A38 X39
        0.06993612265363  X36 X38 X39
        0.04857319456736  X36 A37 X38 B39
        0.04857319456736  X36 B37 X38 A39
        0.03387125587282  X37 X38 X39
```

## MECI-2 (Closed)

```
26
C   -0.1650232956   0.5199986019  -0.6098533066
H    0.0239726792   0.5457284698  -1.6804300848
C   -1.2961806648   1.5567880350  -0.2977361226
C   -0.4913113317  -0.9174208616  -0.2849800925
H    1.2797712699  -1.3234433570   0.7499854501
C    1.5850080573  -1.9355937579  -1.2707766007
C   -2.7069242833   1.0090631918  -0.4782102455
C   -2.7836193177  -0.3320466285  -0.0784861660
C   -1.7709125930  -1.3516300764  -0.2648537042
H   -3.6364788243  -0.7018443063   0.4902313771
H    0.7758962314   0.7946603199  -0.1376034628
C    0.3394236904  -3.2205788136   0.5019639326
H   -1.1303442377   2.4333294176  -0.9142500751
H   -1.1869092437   1.8884240298   0.7317092164
C    0.6950632899  -1.8267354453  -0.0187008367
H    1.0387425844  -2.3871576466  -2.0934957640
H    2.4465489669  -2.5592301591  -1.0575296258
H    1.9496479268  -0.9693061416  -1.5965922612
H    1.2476577607  -3.7573317294   0.7544084864
H   -0.1885645344  -3.8045996468  -0.2450697195
H   -0.2754774635  -3.1731799584   1.3937272309
C   -2.7813415840   0.4870731604  -2.0588586907
H   -2.0104243388  -2.3769828919  -0.0521283398
H   -3.0801301881   1.4465002737  -2.4694008243
H   -3.5989657990  -0.1909890672  -2.2797368024
H   -1.8871427953   0.1088016742  -2.5429491535
```

State Energies and CI Vectors with determinants (X-closed, A-alpha open, B-beta open)

State 1 energy: -387.78474703713619 [Ha]
```
         0.85203509314393   X36 X37 X38
         0.34544544332788   X36 X37 A38 B39
         0.34544544332788   X36 X37 B38 A39
       - 0.09446148523960   X36 A37 X38 B39
       - 0.09446148523960   X36 B37 X38 A39
       - 0.09074708397959   X36 X38 X39
       - 0.05826278339449   X37 X38 X39
         0.05020156153969   X36 A37 B38 X39
         0.05020156153969   X36 B37 A38 X39
```

State 2 energy: -387.78474693419707 [Ha]
```
         0.60689745790252   X36 X37 A38 B39
         0.60689745790252   X36 X37 B38 A39
       - 0.48517680332621   X36 X37 X38
         0.08886250780664   X36 A37 B38 X39
         0.08886250780664   X36 B37 A38 X39
         0.05449782213901   X36 X38 X39
         0.04990844259847   X36 A37 X38 B39
         0.04990844259847   X36 B37 X38 A39
         0.04096810069100   A36 X37 B38 X39
         0.04096810069100   B36 X37 A38 X39
         0.02671076197524   X37 X38 X39
```

## S₁ Minimum (Pericyclic)

```
26
C    0.5538530544  -0.1313875550   0.2183201195
C    2.0551500639  -0.1684684312   0.3970601200
C   -0.0099454398   0.7235677691  -0.8429098022
C   -1.2861865620   1.5044437786  -0.2559467761
C   -2.2879221361   0.5235195051   0.2494951384
C   -0.3302725461  -0.8393476952   1.0693035767
C    2.5365507179  -0.5695493337   1.7947053506
C    2.6804176678  -1.0813235466  -0.6764059461
C   -1.6543142917  -0.4480196128   1.1389425406
C   -3.0124690564  -0.2360198774  -0.8531555823
H    2.4057076786   0.8430178336   0.2019333788
H    0.7460333177   1.3910548516  -1.2486595290
H   -0.3985381681   0.1230927739  -1.6610721337
H   -1.6841655152   2.1236189642  -1.0527210175
H   -0.9198630018   2.1727667385   0.5196358365
H    0.0659191424  -1.4805282423   1.8349239125
H    2.3337048254  -1.6132511871   2.0107365433
H    2.0717045984   0.0330205727   2.5675432295
H    3.6100997959  -0.4276859427   1.8621897283
H    2.4104775234  -0.7638262021  -1.6771165393
H    2.3532012222  -2.1084950976  -0.5478817095
H    3.7626638613  -1.0611434576  -0.5978904298
H   -2.2239308077  -0.8937531809   1.9449294913
H   -3.4947873133   0.4658202399  -1.5279853807
H   -3.7955762568  -0.8760432724  -0.4522558370
H   -2.3759445952  -0.8897656975  -1.4753945693
```

State Energies and CI Vectors with determinants (X-closed, A-alpha open, B-beta open)

State 1 energy: -387.86568279103687 [Ha]
```
          0.61001631082784  X36 X37 X38
        - 0.53742013546126  X36 X37 A38 B39
        - 0.53742013546126  X36 X37 B38 A39
        - 0.15383851559628  X36 X37 X39
        - 0.09131798053993  X36 A37 B38 X39
        - 0.09131798053993  X36 B37 A38 X39
        - 0.05622676498341  A36 X37 B38 X39
        - 0.05622676498341  B36 X37 A38 X39
        - 0.04147578057299  X36 X38 X39
        - 0.02292618750834  A36 X37 X38 B39
        - 0.02292618750834  B36 X37 X38 A39
```

State 2 energy: -387.80857535550507 [Ha]
```
        - 0.76066957786466  X36 X37 X38
        - 0.41170156418479  X36 X37 A38 B39
        - 0.41170156418479  X36 X37 B38 A39
          0.16547436212424  X36 X38 X39
          0.11742496574367  X36 A37 X38 B39
          0.11742496574367  X36 B37 X38 A39
        - 0.11235555708101  X36 X37 X39
        - 0.07244555223866  X36 A37 B38 X39
        - 0.07244555223866  X36 B37 A38 X39
          0.03698760541675  A36 X37 X38 B39
          0.03698760541675  B36 X37 X38 A39
          0.02391006650548  X37 X38 X39
        - 0.02232255808713  A36 X37 B38 X39
        - 0.02232255808713  B36 X37 A38 X39
```

## MECI-3 (Open)

```
26
C    0.7242154180   -0.2197463690   -0.0638014055
C    2.1870210165   -0.1133100143    0.3533998180
C    0.3431374254    0.4689131116   -1.2238729121
C   -1.3748732502    1.3752187124   -0.0998534492
C   -2.1945424954    0.2406641157    0.0801197621
C   -0.1725737967   -0.9876366249    0.7322325994
C    2.4431003061   -0.1509815512    1.8640506089
C    3.0111312550   -1.1989638860   -0.3612119015
C   -1.5613748165   -0.8733307003    0.6255392951
C   -3.5749952182    0.1695303950   -0.5311169053
H    2.5345087116    0.8494711397   -0.0109571903
H    1.0764834108    1.0837212965   -1.7178189965
H   -0.4291551511    0.0772615589   -1.8578934915
H   -1.6845360375    2.1683357029   -0.7582494923
H   -0.7147063416    1.6777552566    0.6898599574
H    0.2360032418   -1.7202377672    1.4032444878
H    2.2623134523   -1.1329885596    2.2870777101
H    1.8197214790    0.5583026156    2.3968524106
H    3.4815664686    0.0995562294    2.0598610066
H    2.8948648105   -1.1394403287   -1.4370546551
H    2.7053087032   -2.1907230762   -0.0422723048
H    4.0648008203   -1.0815878087   -0.1263833060
H   -2.1727986599   -1.6928089758    0.9754981938
H   -4.1071491980    1.1116829757   -0.4415352108
H   -4.1722080060   -0.5885702697   -0.0326393522
H   -3.5539008755   -0.0866201534   -1.5910422390
```

State Energies and CI Vectors with determinants (X-closed, A-alpha open, B-beta open)

State 1 energy: -387.81979655541488 [Ha]
- 0.63249901097436   X36 X37 X38
  0.53457946026321   X36 X37 A38 B39
  0.53457946026321   X36 X37 B38 A39
  0.11144776024141   X36 X38 X39
- 0.06470570373247   X36 X37 X39
  0.05600335383861   X36 A37 B38 X39
  0.05600335383861   X36 B37 A38 X39
  0.05160411521995   X37 X38 X39
  0.03547183261052   X36 A37 X38 B39
  0.03547183261052   X36 B37 X38 A39

State 2 energy: -387.81979639266785 [Ha]
- 0.74775839892161   X36 X37 X38
- 0.45539732535902   X36 X37 A38 B39
- 0.45539732535902   X36 X37 B38 A39
  0.10177909926983   X36 X38 X39
  0.06670171028697   X37 X38 X39
- 0.05600508932177   X36 X37 X39
- 0.05577488218626   X36 A37 B38 X39
- 0.05577488218626   X36 B37 A38 X39
  0.02740067243060   X36 A37 X38 B39
  0.02740067243060   X36 B37 X38 A39

# Rotamer T

## S$_0$ Minimum

```
26
C    0.7173972169   -0.2194694102   -0.0204341205
C    2.2232161917   -0.3534048099    0.0691644920
C    0.0955660783    1.1097225868   -0.3971337509
C   -1.3140154808    1.2765503187    0.1750631227
C   -2.1589839392    0.0335892820    0.0065600323
C   -0.1086876292   -1.2839768307    0.1259624488
C    2.9212718517    0.0906293920   -1.2252549282
C    2.7986790875    0.3949910558    1.2817738048
C   -1.5481627572   -1.1554214599    0.0108317977
C   -3.6499648808    0.2009561367   -0.0700964396
H    2.4409413645   -1.4088507196    0.2104359801
H    0.7153179304    1.9397205362   -0.0751789672
H    0.0465113169    1.1705662742   -1.4851840829
H   -1.7980886567    2.1313828317   -0.2881698712
H   -1.2511084077    1.5013084930    1.2413258403
H    0.2980992168   -2.2624028762    0.3164338539
H    2.7779416426    1.1504166478   -1.4140266020
H    2.5402910236   -0.4566792085   -2.0816374576
H    3.9906927942   -0.0877725168   -1.1618711247
H    2.3349307614    0.0561122466    2.2023490263
H    2.6428555521    1.4667572563    1.2035419413
H    3.8687667157    0.2257456731    1.3617064436
H   -2.1327817854   -2.0586681239   -0.0506479011
H   -4.0223132628    0.7475084040    0.7944928683
H   -4.1590645848   -0.7553566975   -0.1091154541
H   -3.9362073598    0.7729455184   -0.9498909521
```

State Energies and CI Vectors with determinants (X-closed, A-alpha open, B-beta open)

State 1 energy: -387.96831621487092 [Ha]
```
        0.97980576498994  X36 X37 X38
      - 0.12038787334014  X36 X37 X39
        0.07991521304455  X36 X37 A38 B39
        0.07991521304455  X36 X37 B38 A39
      - 0.07428996175315  X36 X38 X39
      - 0.04598683056649  X36 A37 B38 X39
      - 0.04598683056649  X36 B37 A38 X39
      - 0.03367055373372  A36 X37 B38 X39
      - 0.03367055373372  B36 X37 A38 X39
```

State 2 energy: -387.77545243934946 [Ha]
```
      - 0.67904662689236  X36 X37 A38 B39
      - 0.67904662689236  X36 X37 B38 A39
        0.13099026466863  X36 A37 X38 B39
        0.13099026466863  X36 B37 X38 A39
        0.11955044303558  X36 X37 X39
        0.11313180965101  X36 X37 X38
      - 0.07263728677102  X36 X38 X39
      - 0.05719651102697  X36 A37 B38 X39
      - 0.05719651102697  X36 B37 A38 X39
        0.03183693803268  A36 X37 X38 B39
        0.03183693803268  B36 X37 X38 A39
        0.02556879936013  A36 X37 B38 X39
        0.02556879936013  B36 X37 A38 X39
      - 0.02451605786731  A36 B37 X38 X39
      - 0.02451605786731  B36 A37 X38 X39
```

## MECI-1 (Closed)

```
26
C    0.5457969776   -0.1797573307   -0.4217638679
C    1.2500442845   -0.5079875299   -1.7174682565
C    0.3423478750    1.2333506498    0.0432677183
C   -1.1657736789    1.2428821964    0.3426983404
C   -1.7455981893   -0.0341110161   -0.2430997561
C    0.1825870087   -1.2372423337    0.5310211982
C    0.8214154637    0.3196088200   -2.9339729917
C    2.7663966813   -0.3640514971   -1.4615289307
C   -1.1416179884   -1.2518541758    0.1259943203
C   -3.1160206716    0.0158452655   -0.8577305264
H    1.0503863972   -1.5527225745   -1.9257922556
H    0.9452644140    1.3980304873    0.9338789613
H    0.6060198113    1.9709288164   -0.7056500999
H   -1.6313903563    2.1134904989   -0.1047444896
H   -1.3596482385    1.2868976658    1.4146570204
H    0.7440157386   -2.1519494441    0.5792781138
H    1.0558248385    1.3721289301   -2.8152629161
H   -0.2441027166    0.2251059604   -3.1074362905
H    1.3416488071   -0.0324351579   -3.8201018567
H    3.0915480394   -0.9798683109   -0.6302761316
H    3.0351547980    0.6658538799   -1.2468848436
H    3.3176355606   -0.6727439917   -2.3439862841
H   -1.6973125920   -2.1702914324   -0.0357146582
H   -3.8268372294    0.5054747878   -0.1897313370
H   -3.4953717606   -0.9779220400   -1.0738173113
H   -3.1317182400    0.5789721728   -1.7892859349
```

State Energies and CI Vectors with determinants (X-closed, A-alpha open, B-beta open)

State 1 energy -387.80958737536469 [Ha]
        - 0.69565889775414  X36 X37 A38 B39
        - 0.69565889775414  X36 X37 B38 A39
        - 0.12095689488266  X36 A37 B38 X39
        - 0.12095689488266  X36 B37 A38 X39
          0.03082997786944  X36 A37 X38 B39
          0.03082997786944  X36 B37 X38 A39

State 2 energy: -387.80958729282889 [Ha]
          0.98497434426691  X36 X37 X38
        - 0.09524550840871  X36 A37 X38 B39
        - 0.09524550840871  X36 B37 X38 A39
        - 0.09522828254726  X36 X38 X39
        - 0.04423904514648  X37 X38 X39

## MECI-2 (Closed)

```
26
C   -0.1512859529    0.5553155278   -0.6317841851
H    0.0320934451    0.5776649793   -1.7044500708
C   -1.3092445404    1.5637601664   -0.3125231362
C   -0.4444189066   -0.8868608097   -0.3084683475
H    0.3216645442   -2.8366712369   -0.0574441079
C    1.2314551090   -1.5902851568    1.4123155373
C   -2.7112630617    0.9827554503   -0.4619577539
C   -2.7429934751   -0.3608682837   -0.0601570849
C   -1.7132364691   -1.3531000033   -0.2753663449
H   -3.5721600087   -0.7519399667    0.5296414250
H    0.7830907261    0.8604690073   -0.1695785128
C    1.8437713317   -1.7210128642   -1.0487954072
H   -1.1780773222    2.4371834757   -0.9417504996
H   -1.1880027435    1.9115376376    0.7106071827
C    0.7122693265   -1.8246563589   -0.0161868213
H    1.6009418902   -0.5791636230    1.5468758124
H    2.0472443095   -2.2724278250    1.6281245560
H    0.4473748691   -1.7621671371    2.1416783091
H    2.5723429207   -2.5057703624   -0.8726676624
H    2.3623259136   -0.7706648178   -0.9886627940
H    1.4698710708   -1.8375932499   -2.0609152653
C   -2.8046158651    0.4655530196   -2.0418494932
H   -1.9085404819   -2.3915646380   -0.0646761532
H   -3.1327744032    1.4160147062   -2.4505870565
H   -3.6087058678   -0.2330261296   -2.2499165587
H   -1.9092330131    0.1083460476   -2.5407369806
```

State Energies and CI Vectors with determinants (X-closed, A-alpha open, B-beta open)

State 1 energy: -387.78551795783403 [Ha]
```
         0.89647434252866  X36 X37 X38
       - 0.28347056085296  X36 X37 A38 B39
       - 0.28347056085296  X36 X37 B38 A39
       - 0.09580246425420  X36 X38 X39
       - 0.09521393786426  X36 A37 X38 B39
       - 0.09521393786426  X36 B37 X38 A39
       - 0.05760768739351  X37 X38 X39
       - 0.04111540720766  X36 A37 B38 X39
       - 0.04111540720766  X36 B37 A38 X39
       - 0.02415299889159  A36 X37 B38 X39
       - 0.02415299889159  B36 X37 A38 X39
```

State 2 energy: -387.78551782884199 [Ha]
```
       - 0.63832912111681  X36 X37 A38 B39
       - 0.63832912111681  X36 X37 B38 A39
       - 0.39790306181495  X36 X37 X38
       - 0.09189879883014  X36 A37 B38 X39
       - 0.09189879883014  X36 B37 A38 X39
         0.04606868606279  X36 A37 X38 B39
         0.04606868606279  X36 B37 X38 A39
         0.03968477990647  X36 X38 X39
       - 0.03823235869874  A36 X37 B38 X39
       - 0.03823235869874  B36 X37 A38 X39
         0.03190287168786  X37 X38 X39
```

## S$_1$ Minimum (Pericyclic)

```
26
C    0.5414891843   -0.1432495270    0.3015908264
C    2.0119842827   -0.2084758992    0.6501178597
C   -0.0538114048    1.0805296516   -0.2696032502
C   -1.4526524085    1.3734189193    0.4590500225
C   -2.3567012236    0.1931255622    0.3299013416
C   -0.3003546729   -1.2317629095    0.6291119097
C    2.9011520842    0.0293442713   -0.5800716498
C    2.3536960721    0.7716961164    1.7853386601
C   -1.6631624814   -1.0262255882    0.7414099606
C   -2.8905572959   -0.0132937237   -1.0816299133
H    2.2167228097   -1.2107495527    1.0139149976
H    0.6319765415    1.9203646373   -0.2193539072
H   -0.3034124145    0.9236149555   -1.3160007873
H   -1.8767076374    2.2629715755    0.0053800333
H   -1.2254270395    1.6180393712    1.4942896629
H    0.1400829955   -2.1226702260    1.0440628807
H    2.7820074881    1.0338225442   -0.9708370766
H    2.6693894756   -0.6711354423   -1.3758277735
H    3.9447895000   -0.1026955567   -0.3127823734
H    1.7478104313    0.5804590980    2.6646702030
H    2.2000927144    1.8027505021    1.4868697352
H    3.3960641477    0.6607240011    2.0661908360
H   -2.2140419146   -1.8366415027    1.2031547140
H   -3.4065567859    0.8836290461   -1.4133271829
H   -3.6135331073   -0.8251750057   -1.1232278391
H   -2.1303358804   -0.2454235144   -1.8484054793
```

State Energies and CI Vectors with determinants (X-closed, A-alpha open, B-beta open)

State 1 energy: -387.86592449747934 [Ha]
```
         0.59642927676326  X36 X37 X38
        -0.54525711128160  X36 X37 A38 B39
        -0.54525711128160  X36 X37 B38 A39
        -0.15220403063090  X36 X37 X39
        -0.09250667266292  X36 A37 B38 X39
        -0.09250667266292  X36 B37 A38 X39
        -0.05472135689780  A36 X37 B38 X39
        -0.05472135689780  B36 X37 A38 X39
        -0.04082180350445  X36 X38 X39
        -0.02235876086869  A36 X37 X38 B39
        -0.02235876086869  B36 X37 X38 A39
```

State 2 energy: -387.80930189416142 [Ha]
```
        -0.77156349072755  X36 X37 X38
        -0.40296100178720  X36 X37 A38 B39
        -0.40296100178720  X36 X37 B38 A39
         0.16570049954414  X36 X38 X39
         0.11514169665537  X36 A37 X38 B39
         0.11514169665537  X36 B37 X38 A39
        -0.10667598700915  X36 X37 X39
        -0.07247077793629  X36 A37 B38 X39
        -0.07247077793629  X36 B37 A38 X39
         0.03654480608436  A36 X37 X38 B39
         0.03654480608436  B36 X37 X38 A39
         0.02392219397553  X37 X38 X39
        -0.02041449689276  A36 X37 B38 X39
        -0.02041449689276  B36 X37 A38 X39
```

## MECI-3 (Open)

```
26
C     0.6932110148   -0.1473041829   -0.2933693300
C     2.1760205340   -0.3071200186    0.0229849899
C     0.2270731018    1.0725034782   -0.8014048571
C    -1.3952145346    1.0991439579    0.7579434356
C    -2.2043755551    0.0362269960    0.3089191398
C    -0.1420330671   -1.2586927712    0.0177742140
C     3.0775006675    0.0007162704   -1.1792665521
C     2.5656669265    0.5394767542    1.2448090268
C    -1.5341551014   -1.1658835628    0.0900062893
C    -3.6181923170    0.2702031245   -0.1705268196
H     2.3344896248   -1.3457569488    0.2934915219
H     0.9291155261    1.8763161352   -0.9358850411
H    -0.6149428883    1.0920766186   -1.4661518300
H    -1.7459332984    2.1150075824    0.6990635931
H    -0.6643389968    0.9150924968    1.5204915198
H     0.3200579792   -2.2237998781    0.1196765956
H     3.0554810194    1.0517843884   -1.4423977487
H     2.7740122191   -0.5673334960   -2.0524085357
H     4.1052970305   -0.2635826965   -0.9480370032
H     1.9528651100    0.2855078521    2.1035743545
H     2.4529659704    1.6002830070    1.0476509125
H     3.6027519027    0.3566770866    1.5095680760
H    -2.1155711637   -2.0665898869   -0.0468200451
H    -4.1888222290    0.8929100411    0.5122756811
H    -4.1453498469   -0.6740966229   -0.2675786619
H    -3.6495051253    0.7527845944   -1.1478770285
```

State Energies and CI Vectors with determinants (X-closed, A-alpha open, B-beta open)

State 1 energy: -387.81996898193006 [Ha]
```
          0.77690065026636  X36 X37 X38
        - 0.42970802317920  X36 X37 A38 B39
        - 0.42970802317920  X36 X37 B38 A39
        - 0.10733689851526  X36 X38 X39
        - 0.06933191453799  X37 X38 X39
          0.06047601438592  X36 X37 X39
        - 0.05200031591456  X36 A37 B38 X39
        - 0.05200031591456  X36 B37 A38 X39
        - 0.02716625446757  X36 A37 X38 B39
        - 0.02716625446757  X36 B37 X38 A39
```

State 2 energy: -387.81996890939456 [Ha]
```
        - 0.59554365483223  X36 X37 X38
        - 0.55567694966999  X36 X37 A38 B39
        - 0.55567694966999  X36 X37 B38 A39
          0.10730476802790  X36 X38 X39
        - 0.06599978169045  X36 X37 X39
        - 0.05747916234159  X36 A37 B38 X39
        - 0.05747916234159  X36 B37 A38 X39
          0.05051663859369  X37 X38 X39
          0.03577657249751  X36 A37 X38 B39
          0.03577657249751  X36 B37 X38 A39
```

|  | S₀/S₁ CI (Open) | | | S₀/S1 CI (Closed) | | |
|---|---|---|---|---|---|---|
| **Isomer** | Total | αTP | cZc-IPMHT | Total | αTP | cZc-IPMHT |
| Total | 100 +/- 0 | 42 +/- 4 | 58 +/- 4 | 0 | - | - |
| Rotamer M | 100 +/- 0 | 40 +/- 8 | 60 +/- 8 | 0 | - | - |
| Rotamer P | 100 +/- 0 | 39 +/- 7 | 61 +/- 7 | 0 | - | - |
| Rotamer T | 100 +/- 0 | 48 +/- 7 | 52 +/- 7 | 0 | - | - |

**Table S2. Computational quantum yield for ground state rotamers through the Opening and closed CIs.** Two types of CIs (Open and Closed) are identified. As shown in the table, all the transitions from the excited state to the ground state undergo the Ring-Opening CIs. The quantum yield for the αTP and product isomers are reflected with bootstrap uncertainties.

| Pair distance | α | | β | | γ | |
|---|---|---|---|---|---|---|
| | Exp. | Simu. | Exp. | Simu. | Exp. | Simu. |
| $t_0$ | 115 (23) | 79 (0.2) | 146 (11) | 146 (1) | 26 (15) | 49 (1.3) |
| $\tau$ | 252 (74) | 213 (0.6) | 186 (35) | 257 (3.5) | 48 (48) | 130 (4.0) |

**Table S3. Fitted parameters of the error function fit of the signal onset time and the width.** As shown in Equ. 13, $t_0$ and $\tau$ are corresponding to the shift (signal onset time) and the width in the error function. The values in the parentheses indicate the uncertainty of the fits by taking 68% confidence. All the values are in units of femtoseconds. The fitted signal can be viewed in Fig. S17 with the signal onset time labeled.

**REFERENCES**

1. Weathersby, S. P. *et al.* Mega-electron-volt ultrafast electron diffraction at SLAC National Accelerator Laboratory. *Rev. Sci. Instrum.* **86**, 073702 (2015).

2. Shen, X. *et al.* Femtosecond gas-phase mega-electron-volt ultrafast electron diffraction. *Struct. Dyn.* **6**, 054305 (2019).

3. Yang, J. *et al.* Imaging $CF_3I$ conical intersection and photodissociation dynamics with ultrafast electron diffraction. *Science* **361**, 64–67 (2018).

4. Wolf, T. J. A. *et al.* The photochemical ring-opening of 1,3-cyclohexadiene imaged by ultrafast electron diffraction. *Nat. Chem.* **11**, 504–509 (2019).

5. Snyder, J. W., Hohenstein, E. G., Luehr, N. & Martínez, T. J. An atomic orbital-based formulation of analytical gradients and nonadiabatic coupling vector elements for the state-averaged complete active space self-consistent field method on graphical processing units. *J. Chem. Phys.* **143**, 154107 (2015).

6. Snyder, J. W., Curchod, B. F. E. & Martínez, T. J. GPU-Accelerated State-Averaged Complete Active Space Self-Consistent Field Interfaced with Ab Initio Multiple Spawning Unravels the Photodynamics of Provitamin $D_3$. *J. Phys. Chem. Lett.* **7**, 2444–2449 (2016).

7. Snyder, J. W., Fales, B. S., Hohenstein, E. G., Levine, B. G. & Martínez, T. J. A direct-compatible formulation of the coupled perturbed complete active space self-consistent field equations on graphical processing units. *J. Chem. Phys.* **146**, 174113 (2017).

8. Champenois, E. G. *et al.* Conformer-specific photochemistry imaged in real space and time. *Science* **374**, 178–182 (2021).

9. Ufimtsev, I. S. & Martínez, T. J. Quantum Chemistry on Graphical Processing Units. 1. Strategies for Two-Electron Integral Evaluation. *J. Chem. Theory Comput.* **4**, 222–231 (2008).

10. Ufimtsev, I. S. & Martinez, T. J. Quantum Chemistry on Graphical Processing Units. 2. Direct Self-Consistent-Field Implementation. *J. Chem. Theory Comput.* **5**, 1004–1015 (2009).

11. Ufimtsev, I. S. & Martinez, T. J. Quantum Chemistry on Graphical Processing Units. 3. Analytical Energy Gradients, Geometry Optimization, and First Principles Molecular Dynamics. *J. Chem. Theory Comput.* **5**, 2619–2628 (2009).

12. Ben-Nun, M. & Martínez, T. J. Nonadiabatic molecular dynamics: Validation of the multiple spawning method for a multidimensional problem. *J. Chem. Phys.* **108**, 7244–7257 (1998).

13. Ben-Nun, M., Quenneville, J. & Martínez, T. J. Ab Initio Multiple Spawning: Photochemistry from First Principles Quantum Molecular Dynamics. *J. Phys. Chem. A* **104**, 5161–5175 (2000).

14. Ben-Nun, M. & Martínez, Todd. J. *Ab Initio* Quantum Molecular Dynamics: *Ab Initio* Quantum Molecular Dynamics. in *Advances in Chemical Physics* (eds. Prigogine, I. & Rice, S. A.) 439–512 (John Wiley & Sons, Inc., 2002). doi:10.1002/0471264318.ch7.

15. Perdew, J. P., Burke, K. & Ernzerhof, M. Generalized Gradient Approximation Made Simple. *Phys. Rev. Lett.* **77**, 3865–3868 (1996).

16. Yang, J. *et al.* Simultaneous observation of nuclear and electronic dynamics by ultrafast electron diffraction. *Science* **368**, 885–889 (2020).

17. Ihee, H., Goodson, B. M., Srinivasan, R., Lobastov, V. A. & Zewail, A. H. Ultrafast Electron Diffraction and Structural Dynamics: Transient Intermediates in the Elimination Reaction of $C_2F_4I_2$. *J. Phys. Chem. A* **106**, 4087–4103 (2002).

18. Feenstra, J. S., Park, S. T. & Zewail, A. H. Excited state molecular structures and reactions directly determined by ultrafast electron diffraction. *J. Chem. Phys.* **123**, 221104 (2005).

19. Park, S. T., Feenstra, J. S. & Zewail, A. H. Ultrafast electron diffraction: Excited state structures and chemistries of aromatic carbonyls. *J. Chem. Phys.* **124**, 174707 (2006).

20. Salvat, F., Jablonski, A. & Powell, C. J. elsepa—Dirac partial-wave calculation of elastic scattering of electrons and positrons by atoms, positive ions and molecules. *Comput. Phys. Commun.* **165**, 157–190 (2005).

21. Thompson, A. L., Punwong, C. & Martínez, T. J. Optimization of width parameters for quantum dynamics with frozen Gaussian basis sets. *Chem. Phys.* **370**, 70–77 (2010).

22. Makhov, D. V., Glover, W. J., Martinez, T. J. & Shalashilin, D. V. *Ab initio* multiple cloning algorithm for quantum nonadiabatic molecular dynamics. *J. Chem. Phys.* **141**, 054110 (2014).

23. Ben-Nun, M. & Martínez, T. J. Direct Observation of Disrotatory Ring-Opening in Photoexcited Cyclobutene Using ab Initio Molecular Dynamics. *J. Am. Chem. Soc.* **122**, 6299–6300 (2000).